\documentclass[aps,prd,showpacs, amssymb,superscriptaddress, notitlepage,floats,floatfix]{revtex4-1}
\usepackage{graphicx}
\usepackage{amssymb}
\usepackage{amsmath}
\usepackage{amsfonts}

\usepackage{bm}
\usepackage{mathrsfs}
\usepackage[colorlinks=true]{hyperref}
\usepackage[utf8]{inputenc} 
\usepackage{mathptmx}      
\usepackage{color}
\usepackage{multirow}
\usepackage{palatino}
\usepackage{float}
\usepackage{journals}
\usepackage{soul}
\usepackage{calrsfs}
\DeclareMathAlphabet{\pazocal}{OMS}{zplm}{m}{n}

\definecolor{maroon}{cmyk}{0,0.87,0.68,0.32}

\begin{document}
\title{Ultra Compact Stars: Reconstructing the Perturbation Potential}

\author{Sebastian H. V\"olkel }
\email{sebastian.voelkel@uni-tuebingen.de}
\affiliation{Theoretical Astrophysics, IAAT, University of T\"ubingen, Germany}
\author{Kostas D. Kokkotas}
\affiliation{Theoretical Astrophysics, IAAT, University of T\"ubingen, Germany}
\date{\today}
\begin{abstract}
In this work we demonstrate how different semi-classical methods can be combined in a novel way to reconstruct the perturbation potential of ultra compact stars. Besides rather general assumptions, the only specific information entering this approach is the spectrum of the {\it trapped} axial quasi-normal modes. In general it is not possible to find a unique solution for the potential in the inverse problem, but instead a family of potentials producing the same spectrum. Nevertheless, this already determines important properties of the involved potential and can be used to rule out many candidate models. A unique solution was found based on the additional natural assumption that the exterior part ($r \gtrsim 3\,M$) is described by the Regge-Wheeler potential. This is true in general relativity for any non-rotating spherically symmetric object. This technique can be potentially applied for the study of deviations from general relativity.
The methods we demonstrate are easy to implement and rather general, therefore we expect them also to be interesting for other fields where inverse spectrum problems are studied, e.g. quantum physics and molecular spectroscopy.
\end{abstract}


\maketitle
%
\section{Introduction}
With the repeated detection of gravitational waves from presumably binary black hole mergers by LIGO \cite{LIGO1,PhysRevLett.116.241103,PhysRevLett.118.221101} and its confirmation of different predictions of general relativity \cite{LIGO2}, it might seems clear that black holes exist in nature and exotic alternatives are more unlikely to exist. However, recent claims of discovered ``echoes'' in LIGO data \cite{2016arXiv161200266A,2017arXiv170103485A} show that there could be room for different kind of alternative black hole models or observable quantum gravitational effects \cite{2016PhRvD..94h4031C,2016PhRvL.116q1101C,Barcelo2017,2017arXiv170407175N,PhysRevD.95.084034,PhysRevD.95.043009,2017JCAP...06..056S}. Although technical parts of the signal analysis have been criticized \cite{2016arXiv161205625A}, a full treatment of the problem and more observations have to be carried out for clarification.
\par
The gravitational wave signature of many alternative ultra compact objects is expected be in agreement with the observed signal during the inspiral phase and interestingly also, at the later phases, for the part of the signal that corresponds to the first ringdown mode. In contrary to black holes, it is predicted that additional structure in the signal shows up at later times, recently often called ``echoes'' \cite{2016arXiv161205625A,2007PhRvD..76b4016D,2016PhRvD..94h4031C,2016PhRvL.116q1101C,2016JCAP...12..043K,2017arXiv170204833P,2017arXiv170405789B,2017arXiv170407175N}, which might be associated with the excitation of {\em trapped modes} of the object \cite{1991RSPSA.432..247C,1994MNRAS.268.1015K,1995RSPSA.451..341K,1996gr.qc.....3024K,1996ApJ...462..855A,PhysRevD.60.024004,1999LRR.....2....2K,1999MNRAS.310..797B,2000PhRvD..62j7504F}. Thus, up to now, it is not possible to rule out such alternative objects based on the available gravitational wave observations and it remains an open problem \cite{2016PhLB..756..350K,2016arXiv161205625A}. For an extended and recent overview about echoes and alternative objects we refer to \cite{2017arXiv170703021C}.
\par
In the case that future detections contain additional structure in the signal, one would obviously be interested in the nature of the object. The conventional way to tackle such a problem is to build a model for the object and check if the predicted signal agrees with observation. This is reasonable and straight forward, but has the drawback that one is limited to the models available and laboriously has to solve the direct problem many times to hopefully fit the observation. In this article, we propose an alternative approach which can be used to study the problem the inverse way.
\par
The inverse problem uses the observable data to recover the properties of the source. More specifically, in this work we show how the trapped axial mode spectrum of the source can be used to reconstruct the potential producing the spectrum. 
Using future gravitational wave observations it should be possible to determine the spectrum from the waveform if the unique features (``echoes'' or ``trapped'' modes) characterizing the specific object are excited following the dominant black-hole-type quasi-normal mode.
\par
It is known that to linear order axial perturbations of spherically symmetric and non-rotating objects reduce to a one-dimensional wave equation for the radial part of the perturbation, called $\Psi(x)$ \cite{1999LRR.....2....2K,1999CQGra..16R.159N,2009CQGra..26p3001B,2011RvMP...83..793K}
\begin{align}\label{waveequation}
\mathrm{\frac{d^2}{d x^2}} \Psi(x) + \left(E_n-V(x)\right)\Psi(x) = 0,
\end{align}
where $\omega_n=\sqrt{E_n}$ are the quasi-normal modes and $V(x)$ an effective potential. The direct problem in this case is to calculate $E_n$ from $V(x)$. The inverse problem is to use the information provided by a known/observed spectrum  $E_n$ in order to re-construct the associated potential $V(x)$. In general, one does not expect to find a unique answer to this problem, but based on rather general and reasonable assumptions, it is possible to simplify the problem significantly and obtain a unique solution in many cases. Here we demonstrate that for ultra compact stars there is a way to extract a unique solution for the potential.

\par
This paper is organized as follows. In section \ref{Inverse Problem} we present the mathematical details of our approach for solving the inverse problem. In section \ref{Application and Results} we demonstrate how the method can be applied to constant density stars and gravastars.
We discuss our findings in section \ref{Discussion} while section \ref{Conclusion} contains our conclusions. We provide more details about our implementation and additional results in the Appendix (section \ref{Appendix}). 
\par
Throughout the paper, we assume $G=c=\hbar=2m=M=1$.
%
\section{The Inverse Problem}\label{Inverse Problem}
Inverse problems arise in various fields of science and engineering and are usually much harder to solve than their direct counterpart. Inverse problems can be ill posed and even for simple problems it might be impossible to find a unique solution. An overview about the inverse problem in the context of this work can be found in \cite{lieb2015studies,MR985100,1980AmJPh..48..432L,2006AmJPh..74..638G}. A classical example for the inverse problem is the question whether one can hear the shape of a drum \cite{10.2307/2313748,Gordon1992}. It turns out that the same spectrum of eigenvalues can be produced by different shapes, what means that the inverse problem in this case is not unique.
\par
\begin{figure}
\includegraphics[width=10cm]{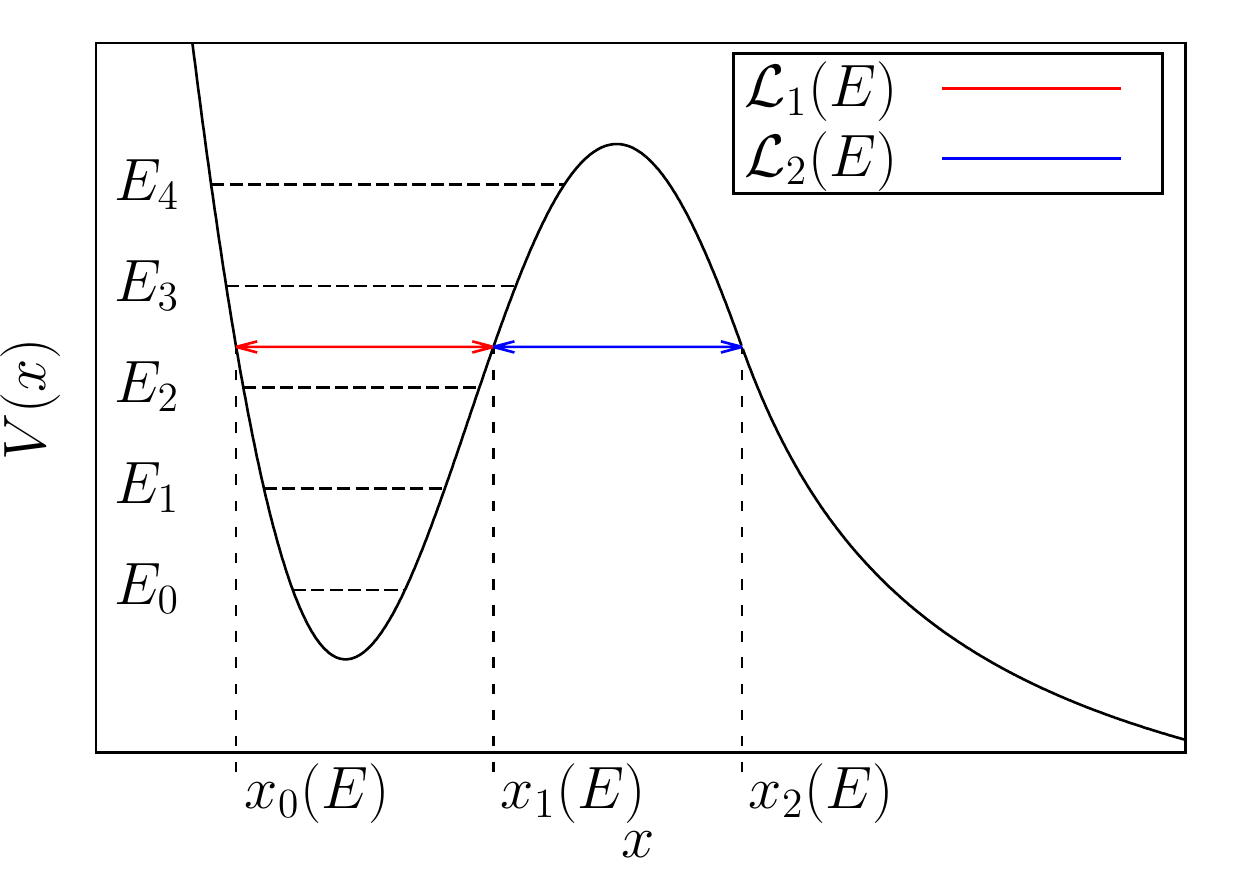}
\caption{Typical potential for quasi-stationary states: shown is the potential $V(x)$, the three classical turning points $(x_0, x_1, x_2)$ for a given state $E_n$ together with the width of the bound region $\pazocal{L}_1(E)$ and potential barrier $\pazocal{L}_2(E)$ at $E=E_n$.\label{WKB-BS}}
\end{figure}

We are interested in the inverse problem for the one-dimensional wave equation \eqref{waveequation} with a typical effective potential for many ultra compact horizonless objects.  The qualitative shape of such a potential can be described by the combination of a bound region next to a potential barrier and is shown in figure \ref{WKB-BS}. The barrier at $r\approx 3\,M$ of the potential corresponds to the photosphere and is the same for black holes and ultra compact stars. The surface of ultra compact stars is somewhere between the potential barrier and the diverging part of the potential, which corresponds to the center of the star. 
For black holes there is no internal barrier and the potential asymptotically tends to zero. 
This difference in the potentials  of the two systems, leads to a completely different quasi-normal mode spectrum.
In \cite{paper1}, hereafter called paper I, we have shown that the widely known Bohr-Sommerfeld methods are useful tools, because they provide a simple framework to obtain useful approximate results and allow for an analytic treatment when considering problems of this kind. Therefore, in this work, we employ such approximate methods and demonstrate how surprisingly precise they can be in solving the inverse problem.
\par
In paper I, we demonstrated that it is possible to treat the direct problem as a combination of a pure bound state and a barrier problem. We adopt the same structure and approach the inverse problem in three steps. First, we calculate the so-called excursion $\pazocal{L}_1(E)$, which is the width of the bound region between the first two turning points $(x_0, x_1)$. Second, we determine the corresponding width of the barrier $\pazocal{L}_2(E)$ between the turning points $(x_1, x_2)$. In the third step we make the additional assumption that the external potential for $r \gtrsim 3\,M$ is described by the Regge-Wheeler potential \cite{1957PhRv..108.1063R}. This assumption provides in principle the third turning point $x_
2$, which together with $\pazocal{L}_1(E)$ and $\pazocal{L}_2(E)$ leads to a unique solution for the potential $V(x)$. 
\par
In literature related to the inverse problem,  one usually finds $E_n$ and not $\omega_n^2$ as definition for the spectrum, thus we adopt this notation. The connection between the trapped quasi-normal modes $\omega_n$ and the spectrum $E_n$ is given by 
\begin{align}
E_n = E_{0n}+iE_{1n} = \Re\left(\omega_n^2\right)+ i \, \Im\left(\omega_n^2\right),
\end{align}
where $E_{0n}$ and $E_{1n}$ are the real and imaginary part of $E_n$. 
In different steps of the method one needs to inter- and extrapolate the spectrum. We provide details for this rather technical issue in the Appendix (section \ref{Notes about the Implementation}).  
%
\subsection{Recovering the Bound Region}
%
The solution to the bound state problem is known for a long time \cite{lieb2015studies,MR985100} and consists of several steps. It has already been applied in different fields, e.g. molecular spectroscopy and quantum algebra \cite{1991JPhA...24L.795B,1992CPL...193..191B,1992JMP....33.2958B}.
\par
The first step is to find $n(E)$ from $E_n$. In a potential of the kind shown in figure \ref{WKB-BS} we expect a finite number of states and a discrete function for $n(E)$. However, in the method one needs to have a continuous function for $n(E)$, because it involves integration. To make it continuous one can interpolate between all bound states and extrapolate to the minimum of the potential.
\par
Starting then from the now approximatively known $n(E)$ one has to calculate the so-called ``inclusion'' $I(E)$ defined as
\begin{align}\label{Inclusion}
I(E) = 2 \int_{E_\text{min}}^{E}\frac{n(E^\prime)+1/2}{\sqrt{E-E^\prime}} \mathrm{d}E^\prime.
\end{align}
Here $E_\text{min}$ is the minimum of the potential. This information is not directly known from the spectrum, but can be approximated from the point where $n(E_\text{min})+1/2$ extrapolates to zero. The inclusion is used as an intermediate function to determine the excursion $\pazocal{L}_1(E)$
\begin{align}\label{Excursion}
\pazocal{L}_1(E) = x_1-x_0 = \frac{\partial }{\partial E} I(E),
\end{align}
which is the width of the bound potential for a given energy $E$ and $(x_0, x_1)$ are the classical turning points. Without further assumptions there are infinitely many WKB equivalent potentials satisfying the above relation and producing the same spectrum. To get a unique solution one has to provide $x_0$ or $x_1$ for one side of the potential. The knowledge of its width $\pazocal{L}_1(E)=x_1-x_0$ then determines the other side. We will use a natural assumption once the barrier region is recovered.
Nevertheless, even the knowledge of $\pazocal{L}_1(E)$ can easily be used to rule out different models, because their corresponding $\pazocal{L}_1(E)$ will in general not agree with the reconstructed one. 
%
\subsection{Recovering the Barrier Region}
%
The semi-classical reconstruction of a one-dimensional potential barrier from the so-called transmission $T(E)$ is known in the literature \cite{1980AmJPh..48..432L,2006AmJPh..74..638G} and we will use the method provided in \cite{2006AmJPh..74..638G}. In our case it will only be possible to determine the width of the barrier $\pazocal{L}_2(E)$ for a given energy
\begin{align}\label{widthbarrier}
 \pazocal{L}_2(E) = x_2-x_1 =  \frac{1}{\pi} \int_{E}^{E_\text{max}} \frac{\left(\text{d}T(E^\prime)/\text{d}E^\prime \right)}{T(E^\prime) \sqrt{E^\prime -E}} \text{d} E^\prime,
\end{align}
here $(x_1, x_2)$ are the classical turning points for the potential barrier and $E_\text{max}$ the maximum of the potential. $E_\text{max}$ is not known from the spectrum but can be extrapolated. $T(E)$ is the transmission and in semi-classical approximation given by
\begin{align}\label{trasmission1}
T(E) = \exp\left(2 i \int_{x_1}^{x_2} \sqrt{E-V(X)} \mathrm{d}x \right).
\end{align}
To reconstruct the potential one needs to know the continuous function $T(E)$, but since we want to limit our knowledge to the trapped quasi-normal modes we only have discrete values for $E_n$. Can the transmission $T(E)$ be determined from the spectrum $E_n$? In our case it is approximatively possible by using that the potential to the left of the barrier is described by a bound region shown in figure \ref{WKB-BS}. Within the Bohr-Sommerfeld framework one can show that the imaginary part of the spectrum for $E_{0n} \gg E_{1n}$ can be approximated with
\begin{align}
E_{1n} = -\frac{1}{2} \exp\left(2 i \int_{x_1}^{x_2} \sqrt{E_{0n}-V(x)} \text{d} x\right)\left(\int_{x_0}^{x_1} \frac{1}{\sqrt{E_{0n}-V(x)}} \text{d} x \right)^{-1},
\end{align}
which contains explicitly the expression for the transmission \eqref{trasmission1} evaluated for $E\approx E_{0n}$, see paper I. The approximation is justified because the imaginary part of trapped modes is typically many orders of magnitude smaller than their real part ($E_{0n} \gg E_{1n}$). The transmission can now be approximated with
\begin{align}\label{transmission2}
T(E_{0n}) = -2 E_{1n} \int_{x_0}^{x_1} \frac{1}{\sqrt{E_{0n}-V(x)}} \text{d} x.
\end{align}
Note that due to the discreteness of $E_{0n}$, one only can calculate only discrete values of the transmission. 
The integral in equation \eqref{transmission2} contains the part of the bound region potential between $x_0$ and $x_1$, which is not known. Fortunately, it can be proven that the integral is equivalent for any potential within the family already known from the reconstruction of $\pazocal{L}_1(E)$, the proof is provided in Appendix (section \ref{EquivalentPotentials}). For simplicity and without loss of generality, one can use the symmetric potential with the turning points $x_{0,1}=\mp \pazocal{L}_1(E)/2$ for integration. The width of the barrier $\pazocal{L}_2(E)$ is then approximatively known after inserting the interpolated transmission into equation \eqref{widthbarrier}.
%
\subsection{A Unique Solution}
%
Without further assumptions one can not find a unique solution for $V(x)$. The reconstructed widths $\pazocal{L}_1(E)$ and $\pazocal{L}_2(E)$ are however very powerful results which can be used to test if a specific model, for any choice of its parameters, can reproduce them. It is trivial to calculate the widths of the bound region and the potential barrier for any given model. This allows to exclude models if the widths differ from the reconstructed ones, but importantly does not prove that the model is correct. There are infinitely many potentials which can be ``tilted'' or ``shifted'', but have the same widths $\pazocal{L}_1(E)$ and $\pazocal{L}_2(E)$. This non-uniqueness disappears if one of the three turning points of the potential is provided.
\par
In general relativity, there is one ``natural'' assumption for non-rotating spherically symmetric objects, that is the  Birkhoff's theorem \cite{Birkhoff}. It states that the external spacetime is described by the Schwarzschild solution. As a result the Regge-Wheeler (RW) potential will uniquely describe the axial 
perturbations of the exterior spacetime.  
Thus for the region on right of the potential barrier i.e. for $r \gtrsim 3\,M$ we will use the RW potential,   defining in a unique way the third turning point $x_2$. It follows that the other two turning points are then given by
\begin{align}
x_1(E) = x_2(E)-\pazocal{L}_2(E), \qquad x_0(E) = x_1(E)-\pazocal{L}_1(E).
\end{align}
Finally, the inversion of $x_0(E)$ and $x_1(E)$ with respect to $E$ determines the unique solution for the reconstructed interior part of the potential $V(x)$. Here $E_\text{min}$ and $E_\text{max}$ correspond to the minimum in the bound region and maximum of the potential barrier, respectively.
\par
There is however one case where we can not use the previously explained assumption to find a unique solution. If the potential in the bound region becomes negative, there is no third corresponding turning point and therefore no unique solution. However, the potential for typical systems like constant density stars and gravastars is positive everywhere, while a negative region would be the signal of  instabilities. Therefore we exclude such cases in the present work. Still, even unstable cases could be  studied by using the same method.
\par
Obviously, the potential which has been reconstructed with approximative methods cannot be exact. There are two different kind of approximations involved. The first one is the general use of semi-classical methods and the second one the specific details of the inter- and extrapolation from the finite number of known states. Only if both kind of approximations are valid one can expect precise results.
%
%
\section{Application and Results}\label{Application and Results}
%
In this section we apply the previously demonstrated methods to two different types of ultra compact objects. Constant density stars are the typical type of  objects for any study of this kind, because they are simple to treat analytically and numerically. Their oscillations have intensively been studied in \cite{1991RSPSA.432..247C,1994MNRAS.268.1015K, 1995RSPSA.451..341K, 1996ApJ...462..855A, 1996gr.qc.....3024K,PhysRevD.60.024004, 2000PhRvD..62j7504F} and in paper I. A second type of ultra compact objects are the more exotic gravastars \cite{2001gr.qc.....9035M}. The values of the trapped mode frequencies  used here, have been derived by a full numerical code discussed in \cite{1994MNRAS.268.1015K} and are tabulated in the Appendix (section \ref{Tables for Quasi-normal Modes}). 
\par
The linear perturbations of both systems, in the non-rotating case, can effectively be described by the one-dimensional wave equation
\begin{equation}
\frac{\mathrm{d}^2}{\mathrm{d} {r^{*}}^2} \Psi(r) + \left(\omega_n^2-V(r) \right) \Psi(r) = 0 ,
\label{eq:waveequation}
\end{equation}
where $\Psi(r)$ is the radial part of the metric perturbation. The wave equation appears in the so-called tortoise coordinate $r^{*}$, which is a function of the normal Schwarzschild coordinate $r$. Its explicit form has to be calculated for every system specifically. For more details we refer to \cite{1999LRR.....2....2K}.
\par
In this work we limit ourselves to the non-rotating case because the axial perturbation equations decouple to a one dimensional wave equation, which does not work for rotating systems. For sure rotation will play an important role for realistic systems, but its treatment within the analytic framework presented here is out of scope of this work. It is reasonable to assume that slow rotation studied in \cite{1992PhRvD..46.4289K,2004PhRvD..70d3003K} will not change the qualitative results. We plan to extend the inverse method also to rotating systems in the future.
%
\subsection{Constant Density Stars}
%
The axial mode potential for constant density stars is given by 
\begin{equation}\label{eq-cs-potential}
V(r) = \frac{e^{2 \nu}}{r^3} \left[l(l+1)r+r^3 (\rho-p(r))-6 M(r) \right] \, ,
\end{equation}
where $l$ is the harmonic index from the expansion of the metric perturbation in spherical tensor harmonics. $\rho$ and $P(r)$ are the density and pressure, respectively. $M(r)$ is the integrated mass function from the center of the star to $r$, while the normalization $4\pi=1$ is also used.
The potential appearing in the wave equation \eqref{eq:waveequation} is expressed in terms of the Schwarzschild coordinate $r$, which is related to the tortoise coordinate $r^{*}$, as follows
\begin{equation}
r^{*} = \int_{0}^{r} e^{-\nu+\mu} \mathrm{d} r \, ,
\end{equation}
where $e^{2\nu}$ and $e^{2\mu}$ are the $g_{00}$ and $g_{11}$ components of the metric tensor $g_{\mu \nu}$. An explicit analytic form can be found in \cite{2000PhRvD..62h4020P}. Constant density stars obey the Buchdahl limit, which sets $R/M=2.25$ as limit for the most compact model.
\par
\begin{figure}[H]
\centering
	\begin{minipage}{5.5cm}
	\includegraphics[width=5.7cm]{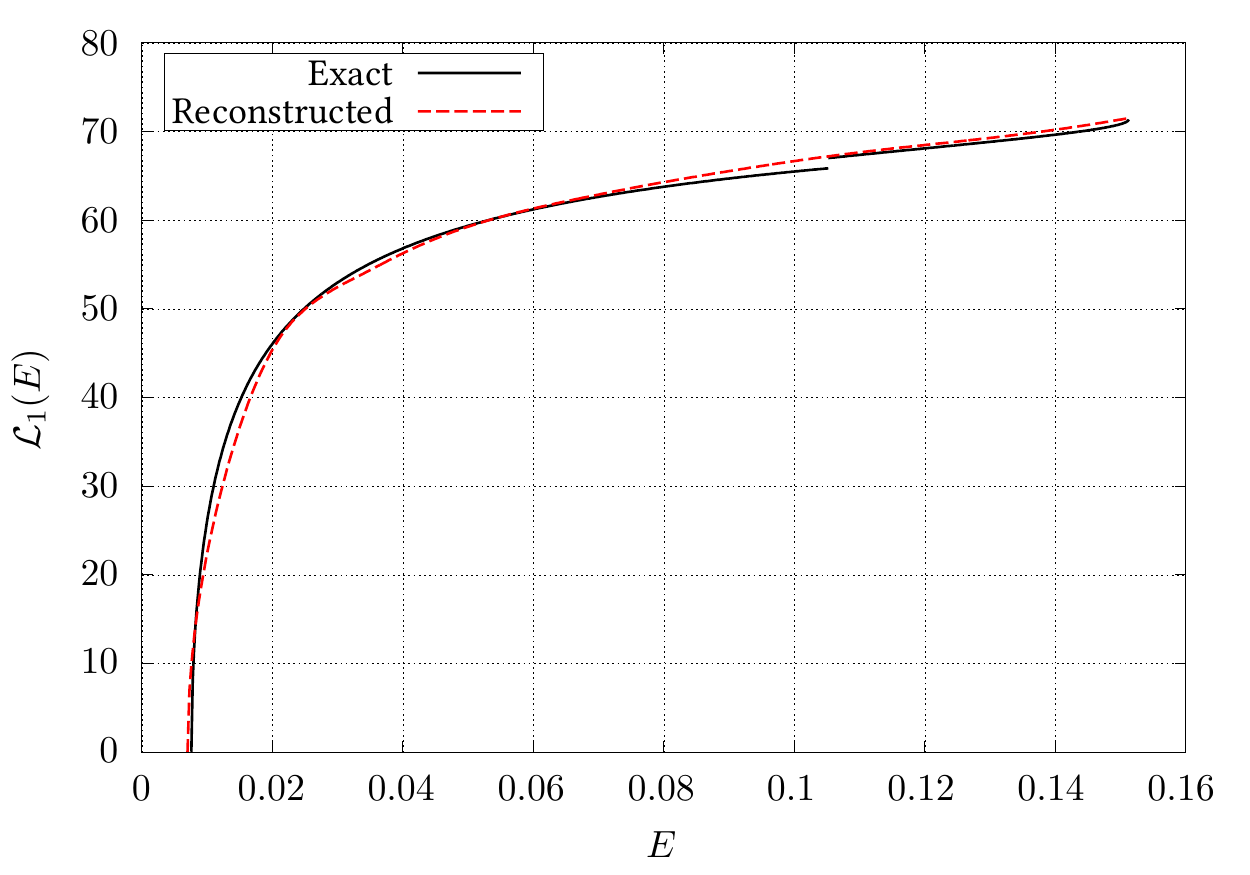}
	\end{minipage}
	\quad
	\begin{minipage}{5.5cm}
	\includegraphics[width=5.7cm]{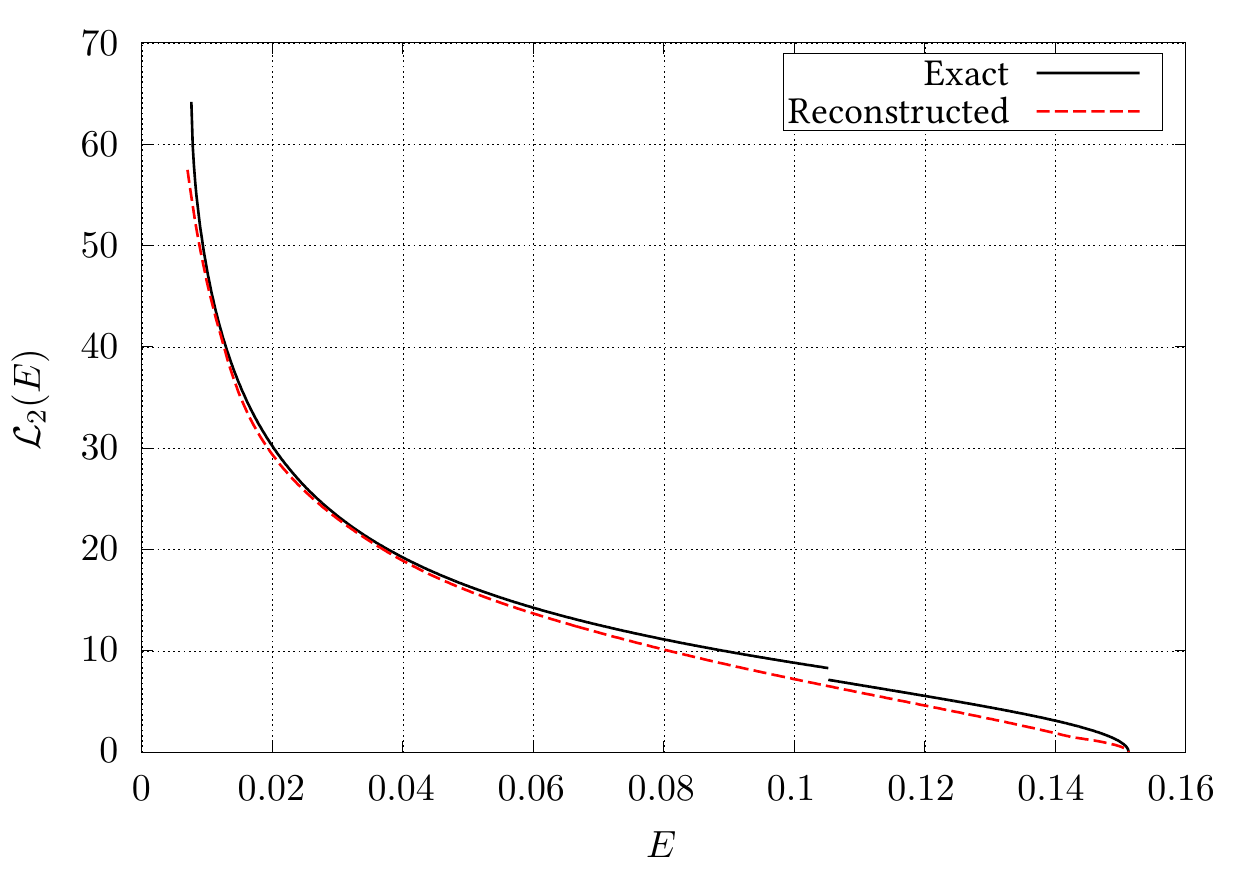}
	\end{minipage}
	\quad
	\begin{minipage}{5.5cm}
	\includegraphics[width=5.7cm]{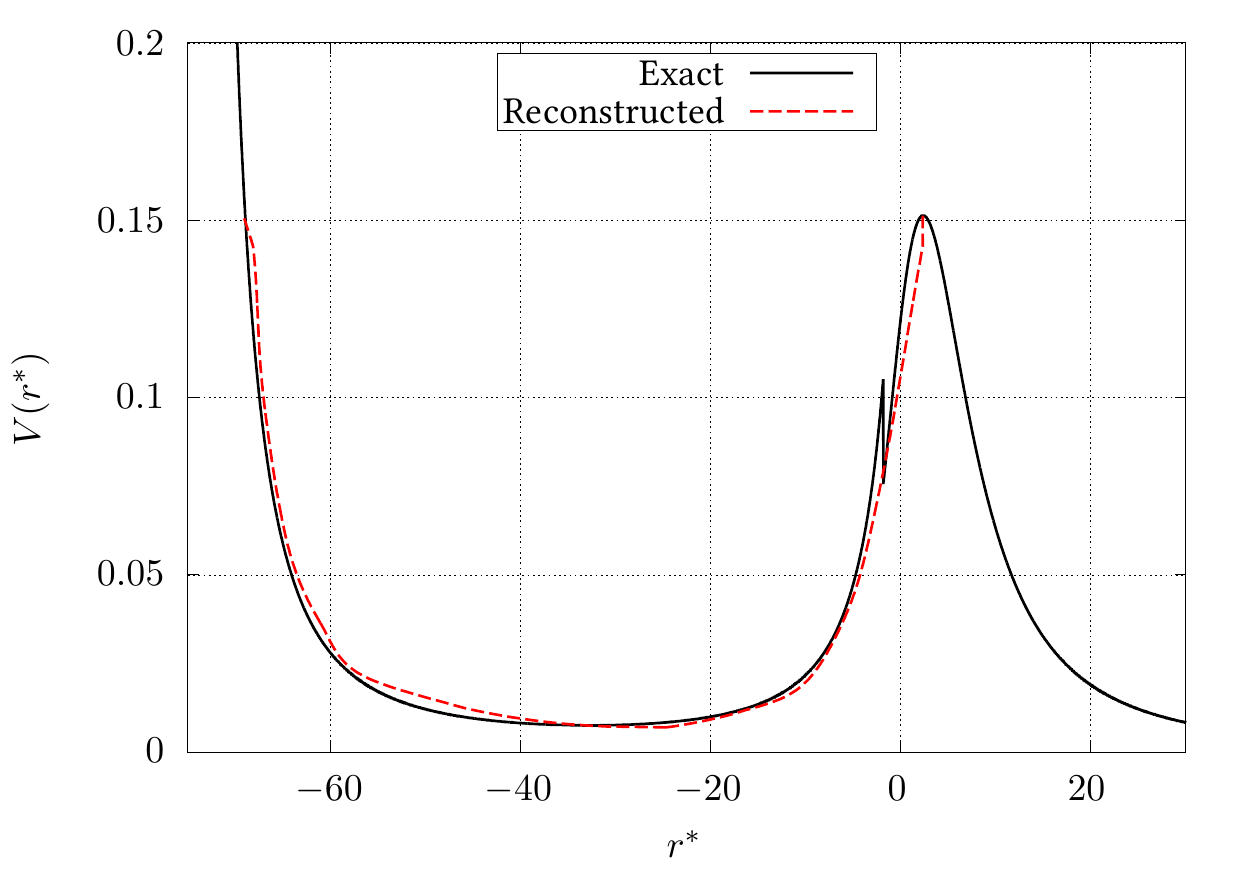}
	\end{minipage}
	\caption{Constant density star with $R/M=2.26$ and $l=2$\label{CS_226_l2}, there are 8 trapped modes. \textbf{Left panel:} width of the bound region $\pazocal{L}_1(E)$ for the exact potential (black) and the reconstructed one (red dashed). \textbf{Central panel:} width of the potential barrier $\pazocal{L}_2(E)$ for the exact potential (black) and the reconstructed one (red dashed). \textbf{Right panel:} exact axial mode potential (black) vs the reconstructed one (red dashed).}
\end{figure}
\begin{figure}[H]
\centering
	\begin{minipage}{5.5cm}
	\includegraphics[width=5.7cm]{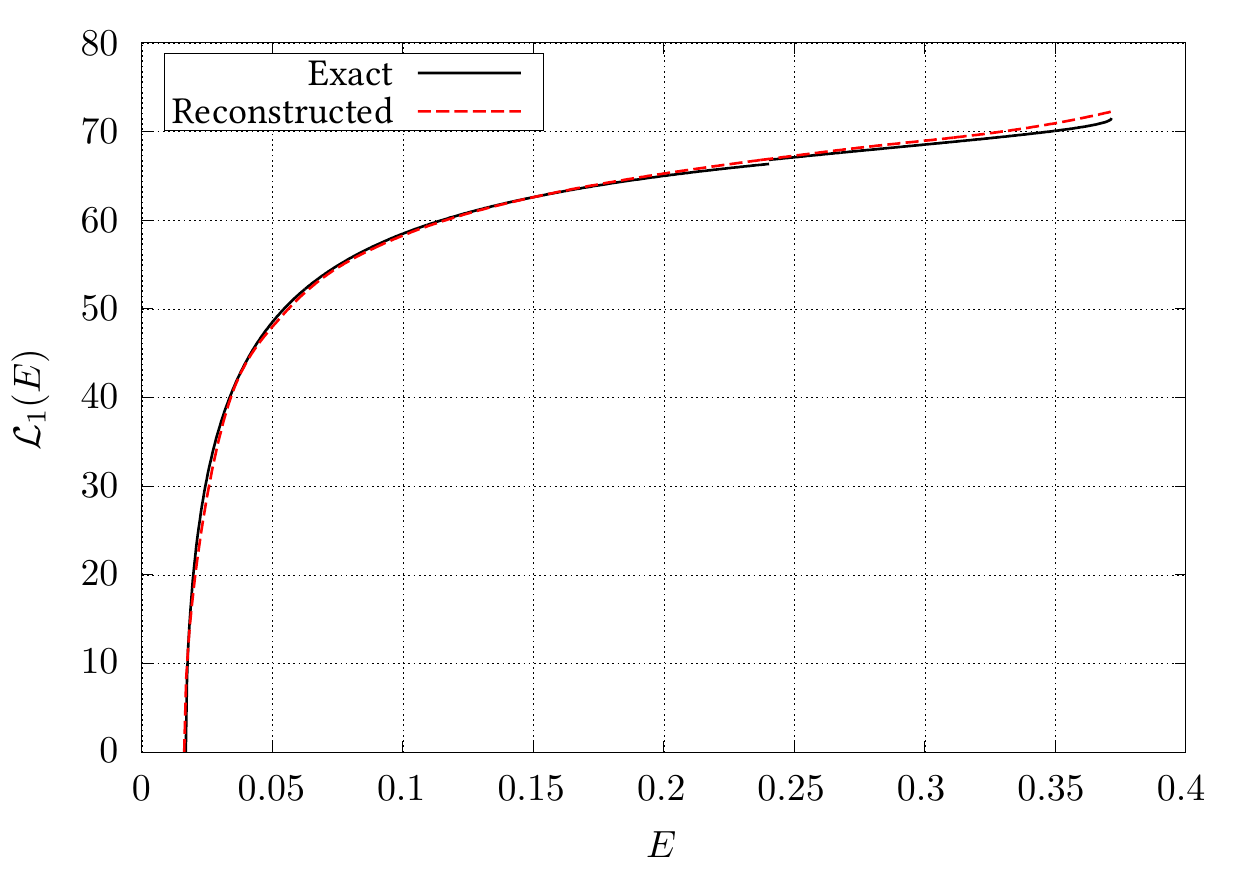}
	\end{minipage}
	\quad
	\begin{minipage}{5.5cm}
	\includegraphics[width=5.7cm]{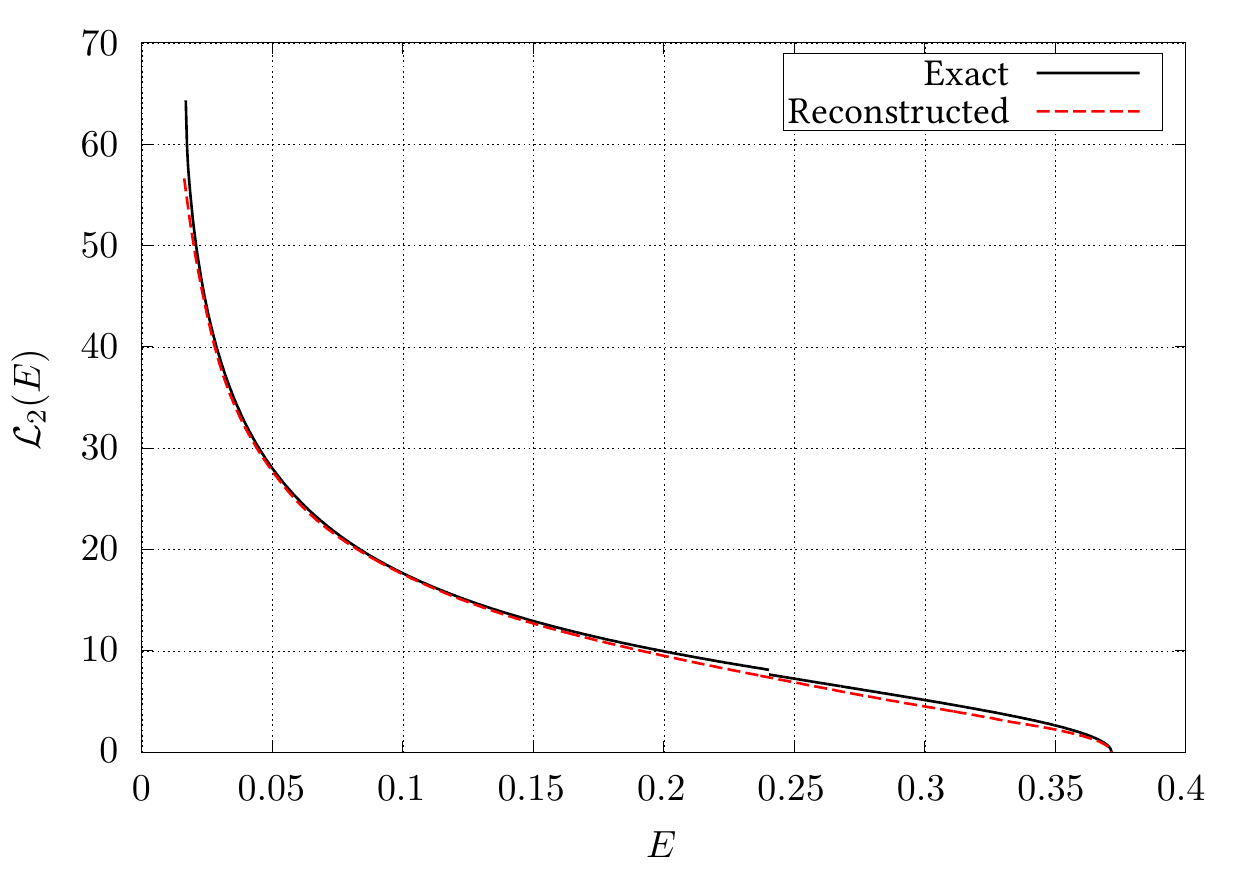}
	\end{minipage}
	\quad
	\begin{minipage}{5.5cm}
	\includegraphics[width=5.7cm]{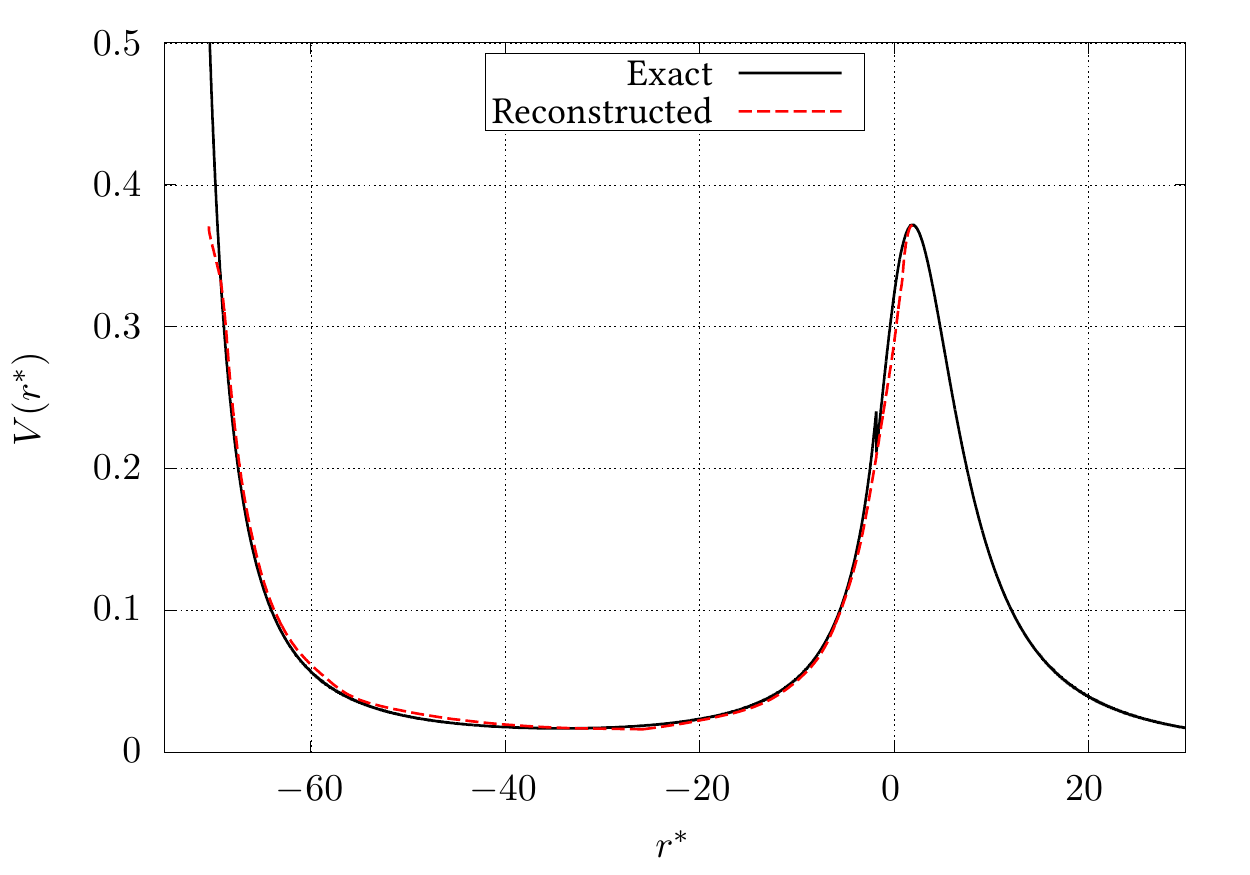}
	\end{minipage}
	\caption{Constant density star with $R/M=2.26$ and $l=3$\label{CS_226_l3}, there are 12 trapped modes. \textbf{Left panel:} width of the bound region $\pazocal{L}_1(E)$ for the exact potential (black) and the reconstructed one (red dashed). \textbf{Central panel:} width of the potential barrier $\pazocal{L}_2(E)$ for the exact potential (black) and the reconstructed one (red dashed). \textbf{Right panel:} exact axial mode potential (black) vs the reconstructed one (red dashed).}
\end{figure}

In figures \ref{CS_226_l2} and \ref{CS_226_l3} we present our results for constant density stars with $R/M=2.26$ and $l=2, \, 3$. More results for $R/M=2.28$ can be found in the Appendix (section \ref{Additional Results}). Each figure consists of three panels and is structured from left to right as follows. The left panel shows the relative width of the bound region $\pazocal{L}_1(E)$ and the central panel shows the width of the potential barrier $\pazocal{L}_2(E)$. The right panel compares the result for the reconstructed potential with the exact potential. In every panel the black solid curve describes the exact function, while the red dashed curve shows our reconstructed result. In the caption we provide the total number of trapped modes that exist in the specific potential and have been used for the reconstruction.
%
\subsection{Gravastars}
%
In contrast to constant density stars, gravastars refer to a wider class of exotic stars. There are models with different layers, where the interior consists of a de Sitter condensate with $p=-\rho$. In this work we study the simplified thin shell model, because our primary interest is to show the applicability of our methods to the inverse problem. The thin shell model assumes an infinitely thin shell at radius $a$, which separates the interior and exterior spacetime. For comprehensive work about the physics of gravastars and their perturbations we refer to \cite{2001gr.qc.....9035M, 2004CQGra..21.1135V, 2007CQGra..24.4191C, 2009PhRvD..80l4047P, 1742-6596-222-1-012032, 2014PhRvD..90d4069C, 2016PhRvD..94h4016C}.
\par
The axial mode potential for thin shell gravastars is given by
\begin{equation}
\label{eq-gs-potential}
V(r) = \left(1-\frac{8 \pi \rho}{3} r^2\right) \frac{l(l+1)}{r^2}.
\end{equation}
Here $l$ is again the harmonic index and $\rho$ the density. This model corresponds to the specific equation of state in the shell with zero surface density. The wave equation is written again in terms of the tortoise coordinate $r^{*}$, which in the interior region is given by the relation
%
\begin{equation}
r^* = \sqrt{\frac{3}{8 \pi \rho}} \text{arctanh}\left[ \left(\frac{8 \pi \rho r^2}{3} \right)^{1/2}\right] + C \, .
\label{gravastar-rtortoise}
\end{equation}
%
Here $C$ is a constant of integration, chosen by demanding that  $r^*$  continuously matched with the usual exterior Schwarzschild tortoise coordinate
\begin{equation}
C = a +  2 M \ln\left(\frac{a}{2M}-1 \right) - \sqrt{\frac{3}{8 \pi \rho}} \text{arctanh} \left[ \left(\frac{8 \pi \rho a^2}{3} \right)^{1/2}\right] ,
\end{equation}
where $a$ stands for the radius of the gravastar. The models were parametrized with respect to  $\mu=M/a$ i.e. the compactness.
\par
\begin{figure}[H]
\centering
	\begin{minipage}{5.5cm}
	\includegraphics[width=5.7cm]{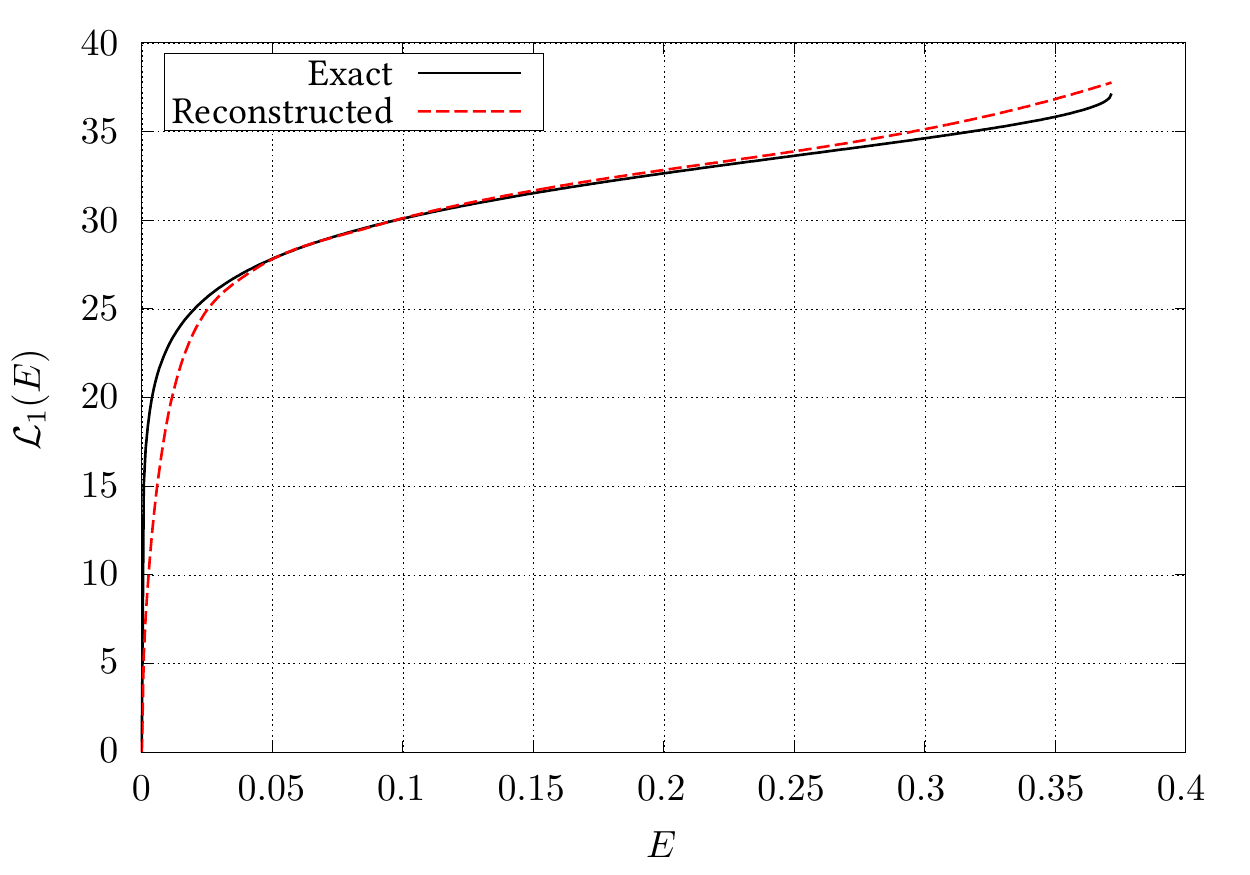}
	\end{minipage}
	\quad
	\begin{minipage}{5.5cm}
	\includegraphics[width=5.7cm]{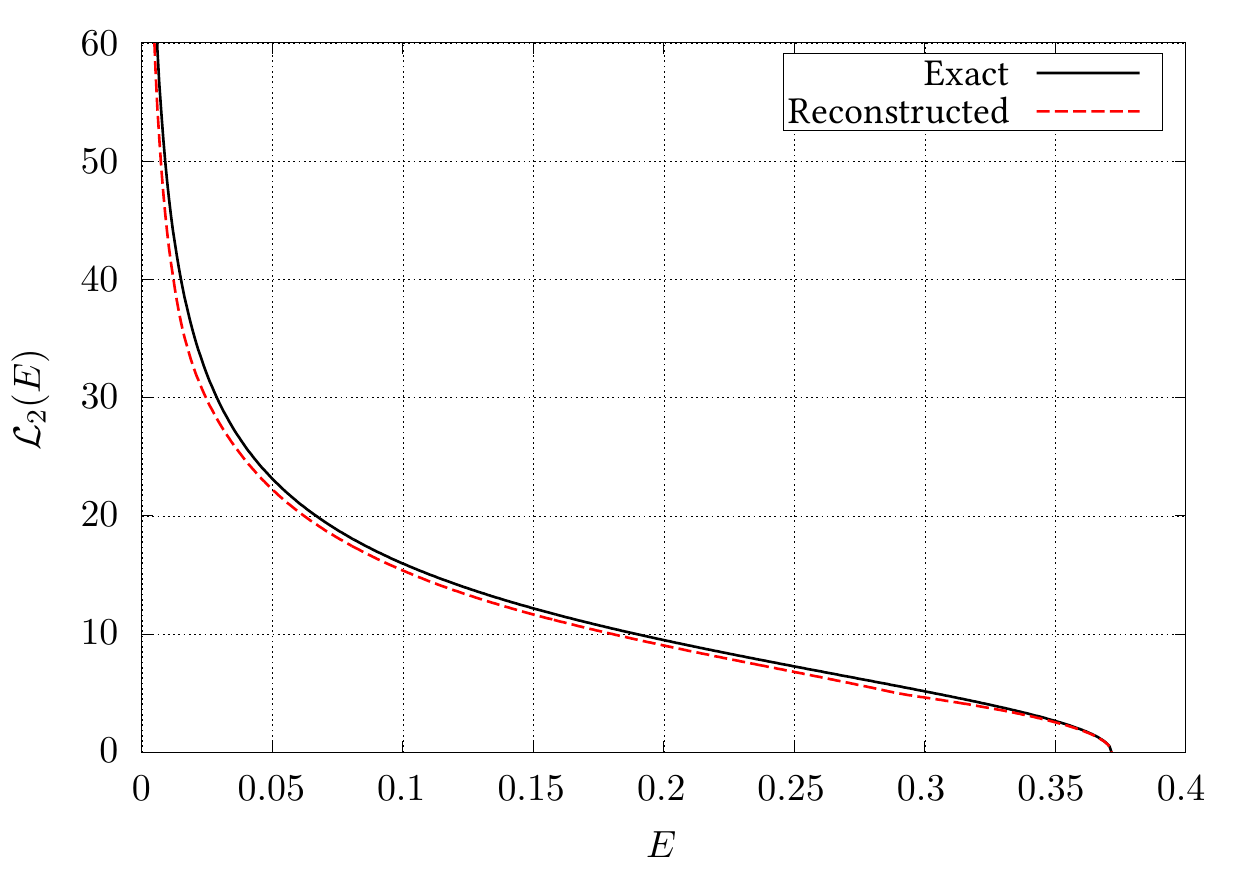}
	\end{minipage}
	\quad
	\begin{minipage}{5.5cm}
	\includegraphics[width=5.7cm]{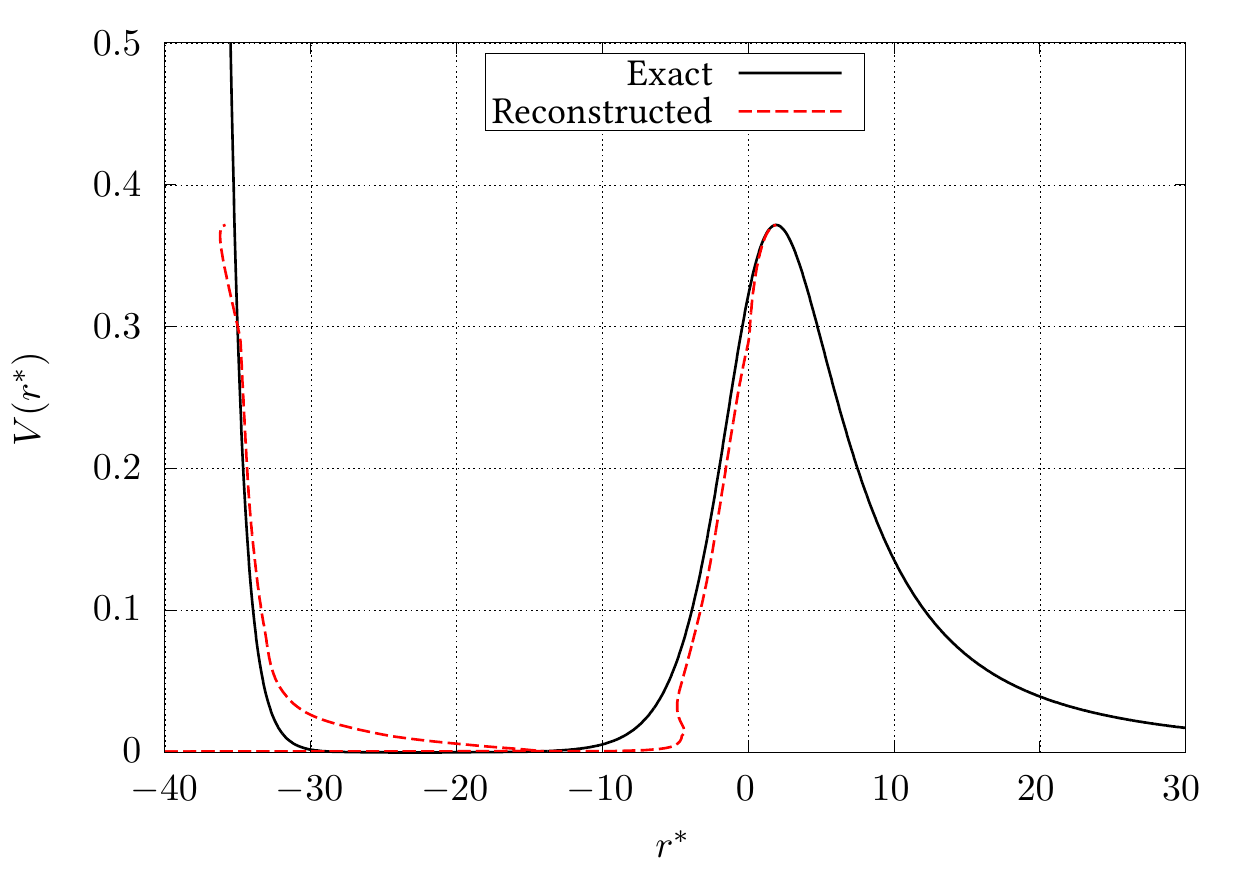}
	\end{minipage}
	\caption{Gravastar with $\mu=0.499999$ and $l=3$\label{GS_499999_l3}, there are 6 trapped modes. \textbf{Left panel:} width of the bound region $\pazocal{L}_1(E)$ for the exact potential (black) and the reconstructed one (red dashed). \textbf{Central panel:} width of the potential barrier $\pazocal{L}_2(E)$ for the exact potential (black) and the reconstructed one (red dashed). \textbf{Right panel:} exact axial mode potential (black) vs the reconstructed one (red dashed).}
\end{figure}

We present results for gravastars with $\mu = 0.499999$ in figure \ref{GS_499999_l3} and provide additional results for less compact gravastars with $\mu = 0.49997$ and $\mu = 0.49999$ in the Appendix (section \ref{Additional Results}). We have chosen results  for $l=3$, since the number of trapped modes for $l=2$ is typically small and the deviations between the true and the reconstructed potential can be large. The appearance of unphysical deformations in the reconstructed potentials for small $E$ is discussed in section \ref{Discussion}.
%
%
\section{Discussion}\label{Discussion}
Here we discuss the results presented in section \ref{Application and Results}. 
In general we find that the agreement between exact and recovered widths, both for $\pazocal{L}_1(E)$ and $\pazocal{L}_2(E)$, is quite good and significantly improves with the number of existing trapped modes in the potential. This connection is natural since the more modes are available, the more accurate become the
interpolations and extrapolations used for deriving the basic ingredients of the method, which are the spectrum
$n(E)$ and the transmission $T(E)$. Because the method is based on WKB theory and for practical purposes has to use inter- and extrapolation, one can not clearly define a minimum number of necessary trapped modes for a desired accuracy of the final result. Our results provide a good estimate what accuracy can be expected for different numbers of existing trapped modes.

Since the height of the potential barrier increases strongly with rising $l$, also the number of existing trapped modes increases. From this one can conclude that the method becomes more precise for large values of $l$, which corresponds to the eikonal limit.

The width of the barrier $\pazocal{L}_2(E)$ seems to be systematically underestimated. This tendency is noticeable if the number of trapped modes is small and becomes less visible otherwise. By studying the direct problem with semi-classical methods in paper I, we found that the imaginary parts of the trapped modes are systematically underestimated for the higher order modes. Thus we make the guess that the systematic deviation is related to the semi-classical methods themselves and does not originate from the details of inter- and extrapolation. 

Comparing the results for the two different types of objects we studied, one finds that the results for constant density stars are significantly more precise than for gravastars. We want to make three comments here. First, the number of trapped modes for the specific gravastar models is smaller and the inter- and extrapolation therefore less precise. Second, from the calculation of the trapped modes for gravastars in paper I, we found that the standard semi-classical result underestimates the fundamental mode. This effect is visible in the underestimation of $\pazocal{L}_1(E)$ for small $E$. Third, the minimum of the gravastar potential is exponentially small, making it hard to be extrapolated.  
This affects the width of the potential barrier $\pazocal{L}_2(E)$. Since it becomes very large for small $E$, the absolute error between the exact width and the reconstructed one grows. The reconstructed potential is sensitive to this absolute error, even if the relative error for the width $\pazocal{L}_2(E)$ is small. This is because the bound region in these cases is much smaller than the width of the barrier. This effect is visible in the unphysical ``overhanging cliffs'' in the reconstructed potential and should be disregarded as artifact for small $E$.

Overall the combined method gives a very good and simple overall estimation of the potential as long as there are enough trapped modes. Since it is intrinsically based on semi-classical methods, one can not expect to resolve any ``fine structure'' in the potential, which is smaller than the associated spacing of the trapped modes. Especially any conclusions for regions of the potential below the first trapped mode have to be drawn carefully. In general the semi-classical methods seem to be reliable for systems with many trapped modes, but one should be careful for the form of the potential near $E_\text{min}$ or in the cases that only a small number of modes is available.  
%
%
\section{Conclusion}\label{Conclusion}
In this work we demonstrated for the first time how the inverse problem for a wide class of ultra compact objects can be solved by using the information provided by the trapped axial quasi-normal modes and the knowledge of the form of the exterior spacetime. 
The inverse problem is very interesting because it allows a mostly model independent study of the properties of the source. 
Our approach is based on a novel combination of two already known semi-classical methods \cite{lieb2015studies,MR985100,1980AmJPh..48..432L,2006AmJPh..74..638G} with  Birkhoff's theorem. We were able to approximatively reproduce the axial mode potential between the minimum and maximum of the quasi-bound region, by using the trapped modes as the only theoretically observable input. The semi-classical methods allow to reconstruct the relative widths of the bound region $\pazocal{L}_1(E)$ and the potential barrier $\pazocal{L}_2(E)$, while Birkhoff's theorem provides a third turning point to determine a unique solution for the potential.
\par
To demonstrate the precision and applicability of the methods we solved the inverse problem for constant density stars and gravastars. 
We find very good agreement as long as the given potential allows for  a reasonable number of trapped modes. 
The method can in principle be used for any one-dimensional wave equation \eqref{waveequation} with a potential of the type shown in figure \ref{WKB-BS}. Therefore it should also find applications in different field of physics where the inverse problem is studied, e.g. molecular spectroscopy or quantum physics. To obtain a unique solution there, one needs to find an alternative to Birkhoff's theorem.
\par
The results of this article were based on the assumption that the complex frequencies are known to pristine accuracy. In reality, this will not be the case. Instead the accuracy in extracting the individual frequencies will depend on the strength of the signal while not all of them will be excited to the same level. This calls for data analysis and information extraction techniques as the ones discussed recently in \cite{2017arXiv170105808Y,2017arXiv170201110M} and should benefit strongly from the future planned ground  (Einstein Telescope \cite{2012CQGra..29l4013S}) and space-based (LISA \cite{PhysRevD.73.064030,PhysRevLett.117.101102}) detectors. A first analysis on how precise such parameters could be extracted from future observations has recently been done in \cite{paper3}.
\par
A natural  extension will be to incorporate the  effects of rotation and the associated ``echoes'', as was recently done in \cite{2017arXiv170303696M,2017arXiv170405856H}. It will be interesting to examine whether the technique presented here can be used to extract information about the rotational state of the star. Trapped modes are excited also for polar oscillations \cite{1995RSPSA.451..341K,1996ApJ...462..855A,2009PhRvD..80l4047P,1742-6596-222-1-012032}, the difficulty in this case is that the spectrum is richer since fluid modes are also excited. Thus the polar oscillation problem cannot be described by a single wave equation of the type (\ref{waveequation}). It is known that polar oscillations will be described by two coupled wave equations \cite{1998PhRvD..58l4012A} and potentially the technique developed here can be applied for this type of problems as well. 
\par
If future gravitational wave detections are able to prove the existence of ``echoes'' or similar structure in the gravitational wave signal \cite{1996gr.qc.....3024K,2000PhRvD..62j7504F,2016PhRvL.116q1101C,2016PhRvD..94h4031C,Barcelo2017}, as recently claimed in \cite{2016arXiv161200266A,2017arXiv170103485A}, this would be a very strong argument that the final object is not an ordinary black hole. Many alternative exotic models of ultra compact objects or possible quantum modifications at the black hole horizon actually predict such a signal and are described by a potential of the kind we studied in this work \cite{2016PhRvD..94h4031C}. Thus, if the additional signal is related to the trapped modes of the object, our method can be a unique and valuable tool in reconstructing the potential and make predictions for the nature of the source. This would be of great importance, because it allows a direct comparison between various phenomenological models proposed for the object and will be a unique tool in understanding its properties.
\acknowledgments
The authors would like to thank Costas Daskaloyannis for useful input in the early stages of the work and Andreas Boden, Roman Konoplya, Vitor Cardoso, Valeria Ferrari, Emanuele Berti, Paolo Pani and Andrea Maselli for useful discussions. The authors appreciate the comments from the anonymous referees that improved the final version of this work.
SV is grateful for the financial support of the Baden-W\"urttemberg Stiftung. This work was partially supported from ``NewCompStar'', Cost Action MP1304.
\bibliography{literatur1}

\begin{thebibliography}{63}%
\makeatletter
\providecommand \@ifxundefined [1]{%
 \@ifx{#1\undefined}
}%
\providecommand \@ifnum [1]{%
 \ifnum #1\expandafter \@firstoftwo
 \else \expandafter \@secondoftwo
 \fi
}%
\providecommand \@ifx [1]{%
 \ifx #1\expandafter \@firstoftwo
 \else \expandafter \@secondoftwo
 \fi
}%
\providecommand \natexlab [1]{#1}%
\providecommand \enquote  [1]{``#1''}%
\providecommand \bibnamefont  [1]{#1}%
\providecommand \bibfnamefont [1]{#1}%
\providecommand \citenamefont [1]{#1}%
\providecommand \href@noop [0]{\@secondoftwo}%
\providecommand \href [0]{\begingroup \@sanitize@url \@href}%
\providecommand \@href[1]{\@@startlink{#1}\@@href}%
\providecommand \@@href[1]{\endgroup#1\@@endlink}%
\providecommand \@sanitize@url [0]{\catcode `\\12\catcode `\$12\catcode
  `\&12\catcode `\#12\catcode `\^12\catcode `\_12\catcode `\%12\relax}%
\providecommand \@@startlink[1]{}%
\providecommand \@@endlink[0]{}%
\providecommand \url  [0]{\begingroup\@sanitize@url \@url }%
\providecommand \@url [1]{\endgroup\@href {#1}{\urlprefix }}%
\providecommand \urlprefix  [0]{URL }%
\providecommand \Eprint [0]{\href }%
\providecommand \doibase [0]{http://dx.doi.org/}%
\providecommand \selectlanguage [0]{\@gobble}%
\providecommand \bibinfo  [0]{\@secondoftwo}%
\providecommand \bibfield  [0]{\@secondoftwo}%
\providecommand \translation [1]{[#1]}%
\providecommand \BibitemOpen [0]{}%
\providecommand \bibitemStop [0]{}%
\providecommand \bibitemNoStop [0]{.\EOS\space}%
\providecommand \EOS [0]{\spacefactor3000\relax}%
\providecommand \BibitemShut  [1]{\csname bibitem#1\endcsname}%
\let\auto@bib@innerbib\@empty
\bibitem [{\citenamefont {{Abbott}}\ \emph {et~al.}(2016)\citenamefont
  {{Abbott}} \emph {et~al.}}]{LIGO1}%
  \BibitemOpen
  \bibfield  {author} {\bibinfo {author} {\bibfnamefont {B.~P.}\ \bibnamefont
  {{Abbott}}} \emph {et~al.} (\bibinfo {collaboration} {LIGO Scientific
  Collaboration and Virgo Collaboration}),\ }\href {\doibase
  10.1103/PhysRevLett.116.061102} {\bibfield  {journal} {\bibinfo  {journal}
  {Phys. Rev. Lett.}\ }\textbf {\bibinfo {volume} {116}},\ \bibinfo {pages}
  {061102} (\bibinfo {year} {2016})}\BibitemShut {NoStop}%
\bibitem [{\citenamefont {Abbott}\ \emph {et~al.}(2016)\citenamefont {Abbott}
  \emph {et~al.}}]{PhysRevLett.116.241103}%
  \BibitemOpen
  \bibfield  {author} {\bibinfo {author} {\bibfnamefont {B.~P.}\ \bibnamefont
  {Abbott}} \emph {et~al.} (\bibinfo {collaboration} {LIGO Scientific
  Collaboration and Virgo Collaboration}),\ }\href {\doibase
  10.1103/PhysRevLett.116.241103} {\bibfield  {journal} {\bibinfo  {journal}
  {Phys. Rev. Lett.}\ }\textbf {\bibinfo {volume} {116}},\ \bibinfo {pages}
  {241103} (\bibinfo {year} {2016})}\BibitemShut {NoStop}%
\bibitem [{\citenamefont {Abbott}\ \emph {et~al.}(2017)\citenamefont {Abbott}
  \emph {et~al.}}]{PhysRevLett.118.221101}%
  \BibitemOpen
  \bibfield  {author} {\bibinfo {author} {\bibfnamefont {B.~P.}\ \bibnamefont
  {Abbott}} \emph {et~al.} (\bibinfo {collaboration} {LIGO Scientific and Virgo
  Collaboration}),\ }\href {\doibase 10.1103/PhysRevLett.118.221101} {\bibfield
   {journal} {\bibinfo  {journal} {Phys. Rev. Lett.}\ }\textbf {\bibinfo
  {volume} {118}},\ \bibinfo {pages} {221101} (\bibinfo {year}
  {2017})}\BibitemShut {NoStop}%
\bibitem [{\citenamefont {{Abbott}}\ \emph {et~al.}(2016)\citenamefont
  {{Abbott}} \emph {et~al.}}]{LIGO2}%
  \BibitemOpen
  \bibfield  {author} {\bibinfo {author} {\bibfnamefont {B.~P.}\ \bibnamefont
  {{Abbott}}} \emph {et~al.} (\bibinfo {collaboration} {LIGO Scientific and
  Virgo Collaborations}),\ }\href {\doibase 10.1103/PhysRevLett.116.221101}
  {\bibfield  {journal} {\bibinfo  {journal} {Phys. Rev. Lett.}\ }\textbf
  {\bibinfo {volume} {116}},\ \bibinfo {pages} {221101} (\bibinfo {year}
  {2016})}\BibitemShut {NoStop}%
\bibitem [{\citenamefont {{Abedi}}\ \emph {et~al.}(2016)\citenamefont
  {{Abedi}}, \citenamefont {{Dykaar}},\ and\ \citenamefont
  {{Afshordi}}}]{2016arXiv161200266A}%
  \BibitemOpen
  \bibfield  {author} {\bibinfo {author} {\bibfnamefont {J.}~\bibnamefont
  {{Abedi}}}, \bibinfo {author} {\bibfnamefont {H.}~\bibnamefont {{Dykaar}}}, \
  and\ \bibinfo {author} {\bibfnamefont {N.}~\bibnamefont {{Afshordi}}},\
  }\href@noop {} {\bibfield  {journal} {\bibinfo  {journal} {ArXiv e-prints}\ }
  (\bibinfo {year} {2016})},\ \Eprint {http://arxiv.org/abs/1612.00266}
  {arXiv:1612.00266 [gr-qc]} \BibitemShut {NoStop}%
\bibitem [{\citenamefont {{Abedi}}\ \emph {et~al.}(2017)\citenamefont
  {{Abedi}}, \citenamefont {{Dykaar}},\ and\ \citenamefont
  {{Afshordi}}}]{2017arXiv170103485A}%
  \BibitemOpen
  \bibfield  {author} {\bibinfo {author} {\bibfnamefont {J.}~\bibnamefont
  {{Abedi}}}, \bibinfo {author} {\bibfnamefont {H.}~\bibnamefont {{Dykaar}}}, \
  and\ \bibinfo {author} {\bibfnamefont {N.}~\bibnamefont {{Afshordi}}},\
  }\href@noop {} {\bibfield  {journal} {\bibinfo  {journal} {ArXiv e-prints}\ }
  (\bibinfo {year} {2017})},\ \Eprint {http://arxiv.org/abs/1701.03485}
  {arXiv:1701.03485 [gr-qc]} \BibitemShut {NoStop}%
\bibitem [{\citenamefont {{Cardoso}}\ \emph
  {et~al.}(2016{\natexlab{a}})\citenamefont {{Cardoso}}, \citenamefont
  {{Hopper}}, \citenamefont {{Macedo}}, \citenamefont {{Palenzuela}},\ and\
  \citenamefont {{Pani}}}]{2016PhRvD..94h4031C}%
  \BibitemOpen
  \bibfield  {author} {\bibinfo {author} {\bibfnamefont {V.}~\bibnamefont
  {{Cardoso}}}, \bibinfo {author} {\bibfnamefont {S.}~\bibnamefont {{Hopper}}},
  \bibinfo {author} {\bibfnamefont {C.~F.~B.}\ \bibnamefont {{Macedo}}},
  \bibinfo {author} {\bibfnamefont {C.}~\bibnamefont {{Palenzuela}}}, \ and\
  \bibinfo {author} {\bibfnamefont {P.}~\bibnamefont {{Pani}}},\ }\href
  {\doibase 10.1103/PhysRevD.94.084031} {\bibfield  {journal} {\bibinfo
  {journal} {\prd}\ }\textbf {\bibinfo {volume} {94}},\ \bibinfo {eid} {084031}
  (\bibinfo {year} {2016}{\natexlab{a}})},\ \Eprint
  {http://arxiv.org/abs/1608.08637} {arXiv:1608.08637 [gr-qc]} \BibitemShut
  {NoStop}%
\bibitem [{\citenamefont {{Cardoso}}\ \emph
  {et~al.}(2016{\natexlab{b}})\citenamefont {{Cardoso}}, \citenamefont
  {{Franzin}},\ and\ \citenamefont {{Pani}}}]{2016PhRvL.116q1101C}%
  \BibitemOpen
  \bibfield  {author} {\bibinfo {author} {\bibfnamefont {V.}~\bibnamefont
  {{Cardoso}}}, \bibinfo {author} {\bibfnamefont {E.}~\bibnamefont
  {{Franzin}}}, \ and\ \bibinfo {author} {\bibfnamefont {P.}~\bibnamefont
  {{Pani}}},\ }\href {\doibase 10.1103/PhysRevLett.116.171101} {\bibfield
  {journal} {\bibinfo  {journal} {Physical Review Letters}\ }\textbf {\bibinfo
  {volume} {116}},\ \bibinfo {eid} {171101} (\bibinfo {year}
  {2016}{\natexlab{b}})},\ \Eprint {http://arxiv.org/abs/1602.07309}
  {arXiv:1602.07309 [gr-qc]} \BibitemShut {NoStop}%
\bibitem [{\citenamefont {Barcel{\'o}}\ \emph {et~al.}(2017)\citenamefont
  {Barcel{\'o}}, \citenamefont {Carballo-Rubio},\ and\ \citenamefont
  {Garay}}]{Barcelo2017}%
  \BibitemOpen
  \bibfield  {author} {\bibinfo {author} {\bibfnamefont {C.}~\bibnamefont
  {Barcel{\'o}}}, \bibinfo {author} {\bibfnamefont {R.}~\bibnamefont
  {Carballo-Rubio}}, \ and\ \bibinfo {author} {\bibfnamefont {L.~J.}\
  \bibnamefont {Garay}},\ }\href {\doibase 10.1007/JHEP05(2017)054} {\bibfield
  {journal} {\bibinfo  {journal} {Journal of High Energy Physics}\ }\textbf
  {\bibinfo {volume} {2017}},\ \bibinfo {pages} {54} (\bibinfo {year}
  {2017})}\BibitemShut {NoStop}%
\bibitem [{\citenamefont {Nakano}\ \emph {et~al.}(2017)\citenamefont {Nakano},
  \citenamefont {Sago}, \citenamefont {Tagoshi},\ and\ \citenamefont
  {Tanaka}}]{2017arXiv170407175N}%
  \BibitemOpen
  \bibfield  {author} {\bibinfo {author} {\bibfnamefont {H.}~\bibnamefont
  {Nakano}}, \bibinfo {author} {\bibfnamefont {N.}~\bibnamefont {Sago}},
  \bibinfo {author} {\bibfnamefont {H.}~\bibnamefont {Tagoshi}}, \ and\
  \bibinfo {author} {\bibfnamefont {T.}~\bibnamefont {Tanaka}},\ }\href
  {\doibase 10.1093/ptep/ptx093} {\bibfield  {journal} {\bibinfo  {journal}
  {Progress of Theoretical and Experimental Physics}\ }\textbf {\bibinfo
  {volume} {2017}},\ \bibinfo {pages} {071E01} (\bibinfo {year}
  {2017})}\BibitemShut {NoStop}%
\bibitem [{\citenamefont {Holdom}\ and\ \citenamefont
  {Ren}(2017)}]{PhysRevD.95.084034}%
  \BibitemOpen
  \bibfield  {author} {\bibinfo {author} {\bibfnamefont {B.}~\bibnamefont
  {Holdom}}\ and\ \bibinfo {author} {\bibfnamefont {J.}~\bibnamefont {Ren}},\
  }\href {\doibase 10.1103/PhysRevD.95.084034} {\bibfield  {journal} {\bibinfo
  {journal} {{Phys. Rev. D}}\ }\textbf {\bibinfo {volume} {95}},\ \bibinfo
  {pages} {084034} (\bibinfo {year} {2017})}\BibitemShut {NoStop}%
\bibitem [{\citenamefont {Novotn\'y}\ \emph {et~al.}(2017)\citenamefont
  {Novotn\'y}, \citenamefont {Hlad\'{\i}k},\ and\ \citenamefont
  {Stuchl\'{\i}k}}]{PhysRevD.95.043009}%
  \BibitemOpen
  \bibfield  {author} {\bibinfo {author} {\bibfnamefont {J.}~\bibnamefont
  {Novotn\'y}}, \bibinfo {author} {\bibfnamefont {J.}~\bibnamefont
  {Hlad\'{\i}k}}, \ and\ \bibinfo {author} {\bibfnamefont {Z.~c.~v.}\
  \bibnamefont {Stuchl\'{\i}k}},\ }\href {\doibase 10.1103/PhysRevD.95.043009}
  {\bibfield  {journal} {\bibinfo  {journal} {Phys. Rev. D}\ }\textbf {\bibinfo
  {volume} {95}},\ \bibinfo {pages} {043009} (\bibinfo {year}
  {2017})}\BibitemShut {NoStop}%
\bibitem [{\citenamefont {{Stuchl{\'{\i}}k}}\ \emph {et~al.}(2017)\citenamefont
  {{Stuchl{\'{\i}}k}}, \citenamefont {{Schee}}, \citenamefont {{Toshmatov}},
  \citenamefont {{Hlad{\'{\i}}k}},\ and\ \citenamefont
  {{Novotn{\'y}}}}]{2017JCAP...06..056S}%
  \BibitemOpen
  \bibfield  {author} {\bibinfo {author} {\bibfnamefont {Z.}~\bibnamefont
  {{Stuchl{\'{\i}}k}}}, \bibinfo {author} {\bibfnamefont {J.}~\bibnamefont
  {{Schee}}}, \bibinfo {author} {\bibfnamefont {B.}~\bibnamefont
  {{Toshmatov}}}, \bibinfo {author} {\bibfnamefont {J.}~\bibnamefont
  {{Hlad{\'{\i}}k}}}, \ and\ \bibinfo {author} {\bibfnamefont {J.}~\bibnamefont
  {{Novotn{\'y}}}},\ }\href {\doibase 10.1088/1475-7516/2017/06/056} {\bibfield
   {journal} {\bibinfo  {journal} {\jcap}\ }\textbf {\bibinfo {volume} {6}},\
  \bibinfo {eid} {056} (\bibinfo {year} {2017})},\ \Eprint
  {http://arxiv.org/abs/1704.07713} {arXiv:1704.07713 [gr-qc]} \BibitemShut
  {NoStop}%
\bibitem [{\citenamefont {{Ashton}}\ \emph {et~al.}(2016)\citenamefont
  {{Ashton}}, \citenamefont {{Birnholtz}}, \citenamefont {{Cabero}},
  \citenamefont {{Capano}}, \citenamefont {{Dent}}, \citenamefont {{Krishnan}},
  \citenamefont {{Meadors}}, \citenamefont {{Nielsen}}, \citenamefont
  {{Nitz}},\ and\ \citenamefont {{Westerweck}}}]{2016arXiv161205625A}%
  \BibitemOpen
  \bibfield  {author} {\bibinfo {author} {\bibfnamefont {G.}~\bibnamefont
  {{Ashton}}}, \bibinfo {author} {\bibfnamefont {O.}~\bibnamefont
  {{Birnholtz}}}, \bibinfo {author} {\bibfnamefont {M.}~\bibnamefont
  {{Cabero}}}, \bibinfo {author} {\bibfnamefont {C.}~\bibnamefont {{Capano}}},
  \bibinfo {author} {\bibfnamefont {T.}~\bibnamefont {{Dent}}}, \bibinfo
  {author} {\bibfnamefont {B.}~\bibnamefont {{Krishnan}}}, \bibinfo {author}
  {\bibfnamefont {G.~D.}\ \bibnamefont {{Meadors}}}, \bibinfo {author}
  {\bibfnamefont {A.~B.}\ \bibnamefont {{Nielsen}}}, \bibinfo {author}
  {\bibfnamefont {A.}~\bibnamefont {{Nitz}}}, \ and\ \bibinfo {author}
  {\bibfnamefont {J.}~\bibnamefont {{Westerweck}}},\ }\href@noop {} {\bibfield
  {journal} {\bibinfo  {journal} {ArXiv e-prints}\ } (\bibinfo {year}
  {2016})},\ \Eprint {http://arxiv.org/abs/1612.05625} {arXiv:1612.05625
  [gr-qc]} \BibitemShut {NoStop}%
\bibitem [{\citenamefont {{Damour}}\ and\ \citenamefont
  {{Solodukhin}}(2007)}]{2007PhRvD..76b4016D}%
  \BibitemOpen
  \bibfield  {author} {\bibinfo {author} {\bibfnamefont {T.}~\bibnamefont
  {{Damour}}}\ and\ \bibinfo {author} {\bibfnamefont {S.~N.}\ \bibnamefont
  {{Solodukhin}}},\ }\href {\doibase 10.1103/PhysRevD.76.024016} {\bibfield
  {journal} {\bibinfo  {journal} {\prd}\ }\textbf {\bibinfo {volume} {76}},\
  \bibinfo {eid} {024016} (\bibinfo {year} {2007})},\ \Eprint
  {http://arxiv.org/abs/0704.2667} {arXiv:0704.2667 [gr-qc]} \BibitemShut
  {NoStop}%
\bibitem [{\citenamefont {{Konoplya}}\ and\ \citenamefont
  {{Zhidenko}}(2016{\natexlab{a}})}]{2016JCAP...12..043K}%
  \BibitemOpen
  \bibfield  {author} {\bibinfo {author} {\bibfnamefont {R.~A.}\ \bibnamefont
  {{Konoplya}}}\ and\ \bibinfo {author} {\bibfnamefont {A.}~\bibnamefont
  {{Zhidenko}}},\ }\href {\doibase 10.1088/1475-7516/2016/12/043} {\bibfield
  {journal} {\bibinfo  {journal} {\jcap}\ }\textbf {\bibinfo {volume} {12}},\
  \bibinfo {eid} {043} (\bibinfo {year} {2016}{\natexlab{a}})},\ \Eprint
  {http://arxiv.org/abs/1606.00517} {arXiv:1606.00517 [gr-qc]} \BibitemShut
  {NoStop}%
\bibitem [{\citenamefont {{Price}}\ and\ \citenamefont
  {{Khanna}}(2017)}]{2017arXiv170204833P}%
  \BibitemOpen
  \bibfield  {author} {\bibinfo {author} {\bibfnamefont {R.}~\bibnamefont
  {{Price}}}\ and\ \bibinfo {author} {\bibfnamefont {G.}~\bibnamefont
  {{Khanna}}},\ }\href@noop {} {\bibfield  {journal} {\bibinfo  {journal}
  {ArXiv e-prints}\ } (\bibinfo {year} {2017})},\ \Eprint
  {http://arxiv.org/abs/1702.04833} {arXiv:1702.04833 [gr-qc]} \BibitemShut
  {NoStop}%
\bibitem [{\citenamefont {{Brustein}}\ \emph {et~al.}(2017)\citenamefont
  {{Brustein}}, \citenamefont {{Medved}},\ and\ \citenamefont
  {{Yagi}}}]{2017arXiv170405789B}%
  \BibitemOpen
  \bibfield  {author} {\bibinfo {author} {\bibfnamefont {R.}~\bibnamefont
  {{Brustein}}}, \bibinfo {author} {\bibfnamefont {A.~J.~M.}\ \bibnamefont
  {{Medved}}}, \ and\ \bibinfo {author} {\bibfnamefont {K.}~\bibnamefont
  {{Yagi}}},\ }\href@noop {} {\bibfield  {journal} {\bibinfo  {journal} {ArXiv
  e-prints}\ } (\bibinfo {year} {2017})},\ \Eprint
  {http://arxiv.org/abs/1704.05789} {arXiv:1704.05789 [gr-qc]} \BibitemShut
  {NoStop}%
\bibitem [{\citenamefont {{Chandrasekhar}}\ and\ \citenamefont
  {{Ferrari}}(1991)}]{1991RSPSA.432..247C}%
  \BibitemOpen
  \bibfield  {author} {\bibinfo {author} {\bibfnamefont {S.}~\bibnamefont
  {{Chandrasekhar}}}\ and\ \bibinfo {author} {\bibfnamefont {V.}~\bibnamefont
  {{Ferrari}}},\ }\href {\doibase 10.1098/rspa.1991.0016} {\bibfield  {journal}
  {\bibinfo  {journal} {Proceedings of the Royal Society of London Series A}\
  }\textbf {\bibinfo {volume} {432}},\ \bibinfo {pages} {247} (\bibinfo {year}
  {1991})}\BibitemShut {NoStop}%
\bibitem [{\citenamefont {{Kokkotas}}(1994)}]{1994MNRAS.268.1015K}%
  \BibitemOpen
  \bibfield  {author} {\bibinfo {author} {\bibfnamefont {K.~D.}\ \bibnamefont
  {{Kokkotas}}},\ }\href {\doibase 10.1093/mnras/268.4.1015} {\bibfield
  {journal} {\bibinfo  {journal} {\mnras}\ }\textbf {\bibinfo {volume} {268}},\
  \bibinfo {pages} {1015} (\bibinfo {year} {1994})}\BibitemShut {NoStop}%
\bibitem [{\citenamefont {{Kojima}}\ \emph {et~al.}(1995)\citenamefont
  {{Kojima}}, \citenamefont {{Andersson}},\ and\ \citenamefont
  {{Kokkotas}}}]{1995RSPSA.451..341K}%
  \BibitemOpen
  \bibfield  {author} {\bibinfo {author} {\bibfnamefont {Y.}~\bibnamefont
  {{Kojima}}}, \bibinfo {author} {\bibfnamefont {N.}~\bibnamefont
  {{Andersson}}}, \ and\ \bibinfo {author} {\bibfnamefont {K.~D.}\ \bibnamefont
  {{Kokkotas}}},\ }\href {\doibase 10.1098/rspa.1995.0129} {\bibfield
  {journal} {\bibinfo  {journal} {Proceedings of the Royal Society of London
  Series A}\ }\textbf {\bibinfo {volume} {451}},\ \bibinfo {pages} {341}
  (\bibinfo {year} {1995})},\ \Eprint {http://arxiv.org/abs/gr-qc/9503012}
  {gr-qc/9503012} \BibitemShut {NoStop}%
\bibitem [{\citenamefont {Kokkotas}(1995)}]{1996gr.qc.....3024K}%
  \BibitemOpen
  \bibfield  {author} {\bibinfo {author} {\bibfnamefont {K.~D.}\ \bibnamefont
  {Kokkotas}},\ }in\ \href@noop {} {\emph {\bibinfo {booktitle} {{Relativistic
  gravitation and gravitational radiation. Proceedings, School of Physics, Les
  Houches, France, September 26-October 6, 1995}}}}\ (\bibinfo {year} {1995})\
  pp.\ \bibinfo {pages} {89--102},\ \Eprint
  {http://arxiv.org/abs/gr-qc/9603024} {arXiv:gr-qc/9603024 [gr-qc]}
  \BibitemShut {NoStop}%
\bibitem [{\citenamefont {{Andersson}}\ \emph {et~al.}(1996)\citenamefont
  {{Andersson}}, \citenamefont {{Kojima}},\ and\ \citenamefont
  {{Kokkotas}}}]{1996ApJ...462..855A}%
  \BibitemOpen
  \bibfield  {author} {\bibinfo {author} {\bibfnamefont {N.}~\bibnamefont
  {{Andersson}}}, \bibinfo {author} {\bibfnamefont {Y.}~\bibnamefont
  {{Kojima}}}, \ and\ \bibinfo {author} {\bibfnamefont {K.~D.}\ \bibnamefont
  {{Kokkotas}}},\ }\href {\doibase 10.1086/177199} {\bibfield  {journal}
  {\bibinfo  {journal} {\apj}\ }\textbf {\bibinfo {volume} {462}},\ \bibinfo
  {pages} {855} (\bibinfo {year} {1996})},\ \Eprint
  {http://arxiv.org/abs/gr-qc/9512048} {gr-qc/9512048} \BibitemShut {NoStop}%
\bibitem [{\citenamefont {Tominaga}\ \emph {et~al.}(1999)\citenamefont
  {Tominaga}, \citenamefont {Saijo},\ and\ \citenamefont
  {Maeda}}]{PhysRevD.60.024004}%
  \BibitemOpen
  \bibfield  {author} {\bibinfo {author} {\bibfnamefont {K.}~\bibnamefont
  {Tominaga}}, \bibinfo {author} {\bibfnamefont {M.}~\bibnamefont {Saijo}}, \
  and\ \bibinfo {author} {\bibfnamefont {K.-i.}\ \bibnamefont {Maeda}},\ }\href
  {\doibase 10.1103/PhysRevD.60.024004} {\bibfield  {journal} {\bibinfo
  {journal} {Phys. Rev. D}\ }\textbf {\bibinfo {volume} {60}},\ \bibinfo
  {pages} {024004} (\bibinfo {year} {1999})}\BibitemShut {NoStop}%
\bibitem [{\citenamefont {{Kokkotas}}\ and\ \citenamefont
  {{Schmidt}}(1999)}]{1999LRR.....2....2K}%
  \BibitemOpen
  \bibfield  {author} {\bibinfo {author} {\bibfnamefont {K.~D.}\ \bibnamefont
  {{Kokkotas}}}\ and\ \bibinfo {author} {\bibfnamefont {B.~G.}\ \bibnamefont
  {{Schmidt}}},\ }\href {\doibase 10.12942/lrr-1999-2} {\bibfield  {journal}
  {\bibinfo  {journal} {Living Reviews in Relativity}\ }\textbf {\bibinfo
  {volume} {2}},\ \bibinfo {eid} {2} (\bibinfo {year} {1999})},\ \Eprint
  {http://arxiv.org/abs/gr-qc/9909058} {gr-qc/9909058} \BibitemShut {NoStop}%
\bibitem [{\citenamefont {{Benhar}}\ \emph {et~al.}(1999)\citenamefont
  {{Benhar}}, \citenamefont {{Berti}},\ and\ \citenamefont
  {{Ferrari}}}]{1999MNRAS.310..797B}%
  \BibitemOpen
  \bibfield  {author} {\bibinfo {author} {\bibfnamefont {O.}~\bibnamefont
  {{Benhar}}}, \bibinfo {author} {\bibfnamefont {E.}~\bibnamefont {{Berti}}}, \
  and\ \bibinfo {author} {\bibfnamefont {V.}~\bibnamefont {{Ferrari}}},\ }\href
  {\doibase 10.1046/j.1365-8711.1999.02983.x} {\bibfield  {journal} {\bibinfo
  {journal} {\mnras}\ }\textbf {\bibinfo {volume} {310}},\ \bibinfo {pages}
  {797} (\bibinfo {year} {1999})},\ \Eprint
  {http://arxiv.org/abs/gr-qc/9901037} {gr-qc/9901037} \BibitemShut {NoStop}%
\bibitem [{\citenamefont {{Ferrari}}\ and\ \citenamefont
  {{Kokkotas}}(2000)}]{2000PhRvD..62j7504F}%
  \BibitemOpen
  \bibfield  {author} {\bibinfo {author} {\bibfnamefont {V.}~\bibnamefont
  {{Ferrari}}}\ and\ \bibinfo {author} {\bibfnamefont {K.~D.}\ \bibnamefont
  {{Kokkotas}}},\ }\href {\doibase 10.1103/PhysRevD.62.107504} {\bibfield
  {journal} {\bibinfo  {journal} {\prd}\ }\textbf {\bibinfo {volume} {62}},\
  \bibinfo {eid} {107504} (\bibinfo {year} {2000})},\ \Eprint
  {http://arxiv.org/abs/gr-qc/0008057} {gr-qc/0008057} \BibitemShut {NoStop}%
\bibitem [{\citenamefont {{Konoplya}}\ and\ \citenamefont
  {{Zhidenko}}(2016{\natexlab{b}})}]{2016PhLB..756..350K}%
  \BibitemOpen
  \bibfield  {author} {\bibinfo {author} {\bibfnamefont {R.}~\bibnamefont
  {{Konoplya}}}\ and\ \bibinfo {author} {\bibfnamefont {A.}~\bibnamefont
  {{Zhidenko}}},\ }\href {\doibase 10.1016/j.physletb.2016.03.044} {\bibfield
  {journal} {\bibinfo  {journal} {Physics Letters B}\ }\textbf {\bibinfo
  {volume} {756}},\ \bibinfo {pages} {350} (\bibinfo {year}
  {2016}{\natexlab{b}})},\ \Eprint {http://arxiv.org/abs/1602.04738}
  {arXiv:1602.04738 [gr-qc]} \BibitemShut {NoStop}%
\bibitem [{\citenamefont {{Cardoso}}\ and\ \citenamefont
  {{Pani}}(2017)}]{2017arXiv170703021C}%
  \BibitemOpen
  \bibfield  {author} {\bibinfo {author} {\bibfnamefont {V.}~\bibnamefont
  {{Cardoso}}}\ and\ \bibinfo {author} {\bibfnamefont {P.}~\bibnamefont
  {{Pani}}},\ }\href@noop {} {\bibfield  {journal} {\bibinfo  {journal} {ArXiv
  e-prints}\ } (\bibinfo {year} {2017})},\ \Eprint
  {http://arxiv.org/abs/1707.03021} {arXiv:1707.03021 [gr-qc]} \BibitemShut
  {NoStop}%
\bibitem [{\citenamefont {{Nollert}}(1999)}]{1999CQGra..16R.159N}%
  \BibitemOpen
  \bibfield  {author} {\bibinfo {author} {\bibfnamefont {H.-P.}\ \bibnamefont
  {{Nollert}}},\ }\href {\doibase 10.1088/0264-9381/16/12/201} {\bibfield
  {journal} {\bibinfo  {journal} {Classical and Quantum Gravity}\ }\textbf
  {\bibinfo {volume} {16}},\ \bibinfo {pages} {R159} (\bibinfo {year}
  {1999})}\BibitemShut {NoStop}%
\bibitem [{\citenamefont {{Berti}}\ \emph {et~al.}(2009)\citenamefont
  {{Berti}}, \citenamefont {{Cardoso}},\ and\ \citenamefont
  {{Starinets}}}]{2009CQGra..26p3001B}%
  \BibitemOpen
  \bibfield  {author} {\bibinfo {author} {\bibfnamefont {E.}~\bibnamefont
  {{Berti}}}, \bibinfo {author} {\bibfnamefont {V.}~\bibnamefont {{Cardoso}}},
  \ and\ \bibinfo {author} {\bibfnamefont {A.~O.}\ \bibnamefont
  {{Starinets}}},\ }\href {\doibase 10.1088/0264-9381/26/16/163001} {\bibfield
  {journal} {\bibinfo  {journal} {Classical and Quantum Gravity}\ }\textbf
  {\bibinfo {volume} {26}},\ \bibinfo {eid} {163001} (\bibinfo {year}
  {2009})},\ \Eprint {http://arxiv.org/abs/0905.2975} {arXiv:0905.2975 [gr-qc]}
  \BibitemShut {NoStop}%
\bibitem [{\citenamefont {{Konoplya}}\ and\ \citenamefont
  {{Zhidenko}}(2011)}]{2011RvMP...83..793K}%
  \BibitemOpen
  \bibfield  {author} {\bibinfo {author} {\bibfnamefont {R.~A.}\ \bibnamefont
  {{Konoplya}}}\ and\ \bibinfo {author} {\bibfnamefont {A.}~\bibnamefont
  {{Zhidenko}}},\ }\href {\doibase 10.1103/RevModPhys.83.793} {\bibfield
  {journal} {\bibinfo  {journal} {Reviews of Modern Physics}\ }\textbf
  {\bibinfo {volume} {83}},\ \bibinfo {pages} {793} (\bibinfo {year} {2011})},\
  \Eprint {http://arxiv.org/abs/1102.4014} {arXiv:1102.4014 [gr-qc]}
  \BibitemShut {NoStop}%
\bibitem [{\citenamefont {Wheeler}(2015)}]{lieb2015studies}%
  \BibitemOpen
  \bibfield  {author} {\bibinfo {author} {\bibfnamefont {J.~A.}\ \bibnamefont
  {Wheeler}},\ }\href@noop {} {\emph {\bibinfo {title} {{Studies in
  Mathematical Physics: Essays in Honor of Valentine Bargmann}}}}\ (\bibinfo
  {publisher} {Princeton University Press},\ \bibinfo {year} {2015})\ pp.\
  \bibinfo {pages} {351--422}\BibitemShut {NoStop}%
\bibitem [{\citenamefont {Chadan}\ and\ \citenamefont
  {Sabatier}(1989)}]{MR985100}%
  \BibitemOpen
  \bibfield  {author} {\bibinfo {author} {\bibfnamefont {K.}~\bibnamefont
  {Chadan}}\ and\ \bibinfo {author} {\bibfnamefont {P.~C.}\ \bibnamefont
  {Sabatier}},\ }\href@noop {} {\emph {\bibinfo {title} {{Inverse problems in
  quantum scattering theory}}}},\ \bibinfo {edition} {2nd}\ ed.,\ Texts and
  Monographs in Physics\ (\bibinfo  {publisher} {Springer-Verlag},\ \bibinfo
  {address} {New York},\ \bibinfo {year} {1989})\BibitemShut {NoStop}%
\bibitem [{\citenamefont {{Lazenby}}\ and\ \citenamefont
  {{Griffiths}}(1980)}]{1980AmJPh..48..432L}%
  \BibitemOpen
  \bibfield  {author} {\bibinfo {author} {\bibfnamefont {J.~C.}\ \bibnamefont
  {{Lazenby}}}\ and\ \bibinfo {author} {\bibfnamefont {D.~J.}\ \bibnamefont
  {{Griffiths}}},\ }\href {\doibase 10.1119/1.11998} {\bibfield  {journal}
  {\bibinfo  {journal} {American Journal of Physics}\ }\textbf {\bibinfo
  {volume} {48}},\ \bibinfo {pages} {432} (\bibinfo {year} {1980})}\BibitemShut
  {NoStop}%
\bibitem [{\citenamefont {{Gandhi}}\ and\ \citenamefont
  {{Efthimiou}}(2006)}]{2006AmJPh..74..638G}%
  \BibitemOpen
  \bibfield  {author} {\bibinfo {author} {\bibfnamefont {S.~C.}\ \bibnamefont
  {{Gandhi}}}\ and\ \bibinfo {author} {\bibfnamefont {C.~J.}\ \bibnamefont
  {{Efthimiou}}},\ }\href {\doibase 10.1119/1.2190683} {\bibfield  {journal}
  {\bibinfo  {journal} {American Journal of Physics}\ }\textbf {\bibinfo
  {volume} {74}},\ \bibinfo {pages} {638} (\bibinfo {year} {2006})},\ \Eprint
  {http://arxiv.org/abs/quant-ph/0503223} {quant-ph/0503223} \BibitemShut
  {NoStop}%
\bibitem [{\citenamefont {Kac}(1966)}]{10.2307/2313748}%
  \BibitemOpen
  \bibfield  {author} {\bibinfo {author} {\bibfnamefont {M.}~\bibnamefont
  {Kac}},\ }\href {http://www.jstor.org/stable/2313748} {\bibfield  {journal}
  {\bibinfo  {journal} {{The American Mathematical Monthly}}\ }\textbf
  {\bibinfo {volume} {73}},\ \bibinfo {pages} {1} (\bibinfo {year}
  {1966})}\BibitemShut {NoStop}%
\bibitem [{\citenamefont {Gordon}\ \emph {et~al.}(1992)\citenamefont {Gordon},
  \citenamefont {Webb},\ and\ \citenamefont {Wolpert}}]{Gordon1992}%
  \BibitemOpen
  \bibfield  {author} {\bibinfo {author} {\bibfnamefont {C.}~\bibnamefont
  {Gordon}}, \bibinfo {author} {\bibfnamefont {D.}~\bibnamefont {Webb}}, \ and\
  \bibinfo {author} {\bibfnamefont {S.}~\bibnamefont {Wolpert}},\ }\href
  {\doibase 10.1007/BF01231320} {\bibfield  {journal} {\bibinfo  {journal}
  {Inventiones mathematicae}\ }\textbf {\bibinfo {volume} {110}},\ \bibinfo
  {pages} {1} (\bibinfo {year} {1992})}\BibitemShut {NoStop}%
\bibitem [{\citenamefont {{V{\"o}lkel}}\ and\ \citenamefont
  {{Kokkotas}}(2017)}]{paper1}%
  \BibitemOpen
  \bibfield  {author} {\bibinfo {author} {\bibfnamefont {S.~H.}\ \bibnamefont
  {{V{\"o}lkel}}}\ and\ \bibinfo {author} {\bibfnamefont {K.~D.}\ \bibnamefont
  {{Kokkotas}}},\ }\href {\doibase 10.1088/1361-6382/aa68cc} {\bibfield
  {journal} {\bibinfo  {journal} {Classical and Quantum Gravity}\ }\textbf
  {\bibinfo {volume} {34}},\ \bibinfo {eid} {125006} (\bibinfo {year}
  {2017})},\ \Eprint {http://arxiv.org/abs/1703.08156} {arXiv:1703.08156
  [gr-qc]} \BibitemShut {NoStop}%
\bibitem [{\citenamefont {{Regge}}\ and\ \citenamefont
  {{Wheeler}}(1957)}]{1957PhRv..108.1063R}%
  \BibitemOpen
  \bibfield  {author} {\bibinfo {author} {\bibfnamefont {T.}~\bibnamefont
  {{Regge}}}\ and\ \bibinfo {author} {\bibfnamefont {J.~A.}\ \bibnamefont
  {{Wheeler}}},\ }\href {\doibase 10.1103/PhysRev.108.1063} {\bibfield
  {journal} {\bibinfo  {journal} {Physical Review}\ }\textbf {\bibinfo {volume}
  {108}},\ \bibinfo {pages} {1063} (\bibinfo {year} {1957})}\BibitemShut
  {NoStop}%
\bibitem [{\citenamefont {{Bonatsos}}\ \emph {et~al.}(1991)\citenamefont
  {{Bonatsos}}, \citenamefont {{Daskaloyannis}},\ and\ \citenamefont
  {{Kokkotas}}}]{1991JPhA...24L.795B}%
  \BibitemOpen
  \bibfield  {author} {\bibinfo {author} {\bibfnamefont {D.}~\bibnamefont
  {{Bonatsos}}}, \bibinfo {author} {\bibfnamefont {C.}~\bibnamefont
  {{Daskaloyannis}}}, \ and\ \bibinfo {author} {\bibfnamefont {K.}~\bibnamefont
  {{Kokkotas}}},\ }\href {\doibase 10.1088/0305-4470/24/15/002} {\bibfield
  {journal} {\bibinfo  {journal} {Journal of Physics A Mathematical General}\
  }\textbf {\bibinfo {volume} {24}},\ \bibinfo {pages} {L795} (\bibinfo {year}
  {1991})}\BibitemShut {NoStop}%
\bibitem [{\citenamefont {{Bonatsos}}\ \emph
  {et~al.}(1992{\natexlab{a}})\citenamefont {{Bonatsos}}, \citenamefont
  {{Daskaloyannis}},\ and\ \citenamefont {{Kokkotas}}}]{1992CPL...193..191B}%
  \BibitemOpen
  \bibfield  {author} {\bibinfo {author} {\bibfnamefont {D.}~\bibnamefont
  {{Bonatsos}}}, \bibinfo {author} {\bibfnamefont {C.}~\bibnamefont
  {{Daskaloyannis}}}, \ and\ \bibinfo {author} {\bibfnamefont {K.}~\bibnamefont
  {{Kokkotas}}},\ }\href {\doibase 10.1016/0009-2614(92)85707-H} {\bibfield
  {journal} {\bibinfo  {journal} {Chemical Physics Letters}\ }\textbf {\bibinfo
  {volume} {193}},\ \bibinfo {pages} {191} (\bibinfo {year}
  {1992}{\natexlab{a}})}\BibitemShut {NoStop}%
\bibitem [{\citenamefont {{Bonatsos}}\ \emph
  {et~al.}(1992{\natexlab{b}})\citenamefont {{Bonatsos}}, \citenamefont
  {{Daskaloyannis}},\ and\ \citenamefont {{Kokkotas}}}]{1992JMP....33.2958B}%
  \BibitemOpen
  \bibfield  {author} {\bibinfo {author} {\bibfnamefont {D.}~\bibnamefont
  {{Bonatsos}}}, \bibinfo {author} {\bibfnamefont {C.}~\bibnamefont
  {{Daskaloyannis}}}, \ and\ \bibinfo {author} {\bibfnamefont {K.}~\bibnamefont
  {{Kokkotas}}},\ }\href {\doibase 10.1063/1.529565} {\bibfield  {journal}
  {\bibinfo  {journal} {Journal of Mathematical Physics}\ }\textbf {\bibinfo
  {volume} {33}},\ \bibinfo {pages} {2958} (\bibinfo {year}
  {1992}{\natexlab{b}})}\BibitemShut {NoStop}%
\bibitem [{\citenamefont {Birkhoff}(1923)}]{Birkhoff}%
  \BibitemOpen
  \bibfield  {author} {\bibinfo {author} {\bibfnamefont {G.~D.}\ \bibnamefont
  {Birkhoff}},\ }\href@noop {} {\emph {\bibinfo {title} {{Relativity and Modern
  Physics}}}}\ (\bibinfo  {publisher} {Cambridge, Massachusetts: Harvard
  University Press},\ \bibinfo {year} {1923})\BibitemShut {NoStop}%
\bibitem [{\citenamefont {{Mazur}}\ and\ \citenamefont
  {{Mottola}}(2001)}]{2001gr.qc.....9035M}%
  \BibitemOpen
  \bibfield  {author} {\bibinfo {author} {\bibfnamefont {P.~O.}\ \bibnamefont
  {{Mazur}}}\ and\ \bibinfo {author} {\bibfnamefont {E.}~\bibnamefont
  {{Mottola}}},\ }\href@noop {} {\bibfield  {journal} {\bibinfo  {journal}
  {ArXiv General Relativity and Quantum Cosmology e-prints}\ } (\bibinfo {year}
  {2001})},\ \Eprint {http://arxiv.org/abs/gr-qc/0109035} {gr-qc/0109035}
  \BibitemShut {NoStop}%
\bibitem [{\citenamefont {{Kojima}}(1992)}]{1992PhRvD..46.4289K}%
  \BibitemOpen
  \bibfield  {author} {\bibinfo {author} {\bibfnamefont {Y.}~\bibnamefont
  {{Kojima}}},\ }\href {\doibase 10.1103/PhysRevD.46.4289} {\bibfield
  {journal} {\bibinfo  {journal} {\prd}\ }\textbf {\bibinfo {volume} {46}},\
  \bibinfo {pages} {4289} (\bibinfo {year} {1992})}\BibitemShut {NoStop}%
\bibitem [{\citenamefont {{Kokkotas}}\ \emph {et~al.}(2004)\citenamefont
  {{Kokkotas}}, \citenamefont {{Ruoff}},\ and\ \citenamefont
  {{Andersson}}}]{2004PhRvD..70d3003K}%
  \BibitemOpen
  \bibfield  {author} {\bibinfo {author} {\bibfnamefont {K.~D.}\ \bibnamefont
  {{Kokkotas}}}, \bibinfo {author} {\bibfnamefont {J.}~\bibnamefont {{Ruoff}}},
  \ and\ \bibinfo {author} {\bibfnamefont {N.}~\bibnamefont {{Andersson}}},\
  }\href {\doibase 10.1103/PhysRevD.70.043003} {\bibfield  {journal} {\bibinfo
  {journal} {\prd}\ }\textbf {\bibinfo {volume} {70}},\ \bibinfo {eid} {043003}
  (\bibinfo {year} {2004})},\ \Eprint {http://arxiv.org/abs/astro-ph/0212429}
  {astro-ph/0212429} \BibitemShut {NoStop}%
\bibitem [{\citenamefont {{Pavlidou}}\ \emph {et~al.}(2000)\citenamefont
  {{Pavlidou}}, \citenamefont {{Tassis}}, \citenamefont {{Baumgarte}},\ and\
  \citenamefont {{Shapiro}}}]{2000PhRvD..62h4020P}%
  \BibitemOpen
  \bibfield  {author} {\bibinfo {author} {\bibfnamefont {V.}~\bibnamefont
  {{Pavlidou}}}, \bibinfo {author} {\bibfnamefont {K.}~\bibnamefont
  {{Tassis}}}, \bibinfo {author} {\bibfnamefont {T.~W.}\ \bibnamefont
  {{Baumgarte}}}, \ and\ \bibinfo {author} {\bibfnamefont {S.~L.}\ \bibnamefont
  {{Shapiro}}},\ }\href {\doibase 10.1103/PhysRevD.62.084020} {\bibfield
  {journal} {\bibinfo  {journal} {\prd}\ }\textbf {\bibinfo {volume} {62}},\
  \bibinfo {eid} {084020} (\bibinfo {year} {2000})},\ \Eprint
  {http://arxiv.org/abs/gr-qc/0007019} {gr-qc/0007019} \BibitemShut {NoStop}%
\bibitem [{\citenamefont {{Visser}}\ and\ \citenamefont
  {{Wiltshire}}(2004)}]{2004CQGra..21.1135V}%
  \BibitemOpen
  \bibfield  {author} {\bibinfo {author} {\bibfnamefont {M.}~\bibnamefont
  {{Visser}}}\ and\ \bibinfo {author} {\bibfnamefont {D.~L.}\ \bibnamefont
  {{Wiltshire}}},\ }\href {\doibase 10.1088/0264-9381/21/4/027} {\bibfield
  {journal} {\bibinfo  {journal} {Classical and Quantum Gravity}\ }\textbf
  {\bibinfo {volume} {21}},\ \bibinfo {pages} {1135} (\bibinfo {year}
  {2004})},\ \Eprint {http://arxiv.org/abs/gr-qc/0310107} {gr-qc/0310107}
  \BibitemShut {NoStop}%
\bibitem [{\citenamefont {{Chirenti}}\ and\ \citenamefont
  {{Rezzolla}}(2007)}]{2007CQGra..24.4191C}%
  \BibitemOpen
  \bibfield  {author} {\bibinfo {author} {\bibfnamefont {C.~B.~M.~H.}\
  \bibnamefont {{Chirenti}}}\ and\ \bibinfo {author} {\bibfnamefont
  {L.}~\bibnamefont {{Rezzolla}}},\ }\href {\doibase
  10.1088/0264-9381/24/16/013} {\bibfield  {journal} {\bibinfo  {journal}
  {Classical and Quantum Gravity}\ }\textbf {\bibinfo {volume} {24}},\ \bibinfo
  {pages} {4191} (\bibinfo {year} {2007})},\ \Eprint
  {http://arxiv.org/abs/0706.1513} {arXiv:0706.1513 [gr-qc]} \BibitemShut
  {NoStop}%
\bibitem [{\citenamefont {{Pani}}\ \emph {et~al.}(2009)\citenamefont {{Pani}},
  \citenamefont {{Berti}}, \citenamefont {{Cardoso}}, \citenamefont {{Chen}},\
  and\ \citenamefont {{Norte}}}]{2009PhRvD..80l4047P}%
  \BibitemOpen
  \bibfield  {author} {\bibinfo {author} {\bibfnamefont {P.}~\bibnamefont
  {{Pani}}}, \bibinfo {author} {\bibfnamefont {E.}~\bibnamefont {{Berti}}},
  \bibinfo {author} {\bibfnamefont {V.}~\bibnamefont {{Cardoso}}}, \bibinfo
  {author} {\bibfnamefont {Y.}~\bibnamefont {{Chen}}}, \ and\ \bibinfo {author}
  {\bibfnamefont {R.}~\bibnamefont {{Norte}}},\ }\href {\doibase
  10.1103/PhysRevD.80.124047} {\bibfield  {journal} {\bibinfo  {journal}
  {\prd}\ }\textbf {\bibinfo {volume} {80}},\ \bibinfo {eid} {124047} (\bibinfo
  {year} {2009})},\ \Eprint {http://arxiv.org/abs/0909.0287} {arXiv:0909.0287
  [gr-qc]} \BibitemShut {NoStop}%
\bibitem [{\citenamefont {{Pani}}\ \emph {et~al.}(2010)\citenamefont {{Pani}},
  \citenamefont {{Berti}}, \citenamefont {{Cardoso}}, \citenamefont {{Chen}},\
  and\ \citenamefont {{Norte}}}]{1742-6596-222-1-012032}%
  \BibitemOpen
  \bibfield  {author} {\bibinfo {author} {\bibfnamefont {P.}~\bibnamefont
  {{Pani}}}, \bibinfo {author} {\bibfnamefont {E.}~\bibnamefont {{Berti}}},
  \bibinfo {author} {\bibfnamefont {V.}~\bibnamefont {{Cardoso}}}, \bibinfo
  {author} {\bibfnamefont {Y.}~\bibnamefont {{Chen}}}, \ and\ \bibinfo {author}
  {\bibfnamefont {R.}~\bibnamefont {{Norte}}},\ }\href
  {http://stacks.iop.org/1742-6596/222/i=1/a=012032} {\bibfield  {journal}
  {\bibinfo  {journal} {Journal of Physics: Conference Series}\ }\textbf
  {\bibinfo {volume} {222}},\ \bibinfo {pages} {012032} (\bibinfo {year}
  {2010})}\BibitemShut {NoStop}%
\bibitem [{\citenamefont {{Cardoso}}\ \emph {et~al.}(2014)\citenamefont
  {{Cardoso}}, \citenamefont {{Crispino}}, \citenamefont {{Macedo}},
  \citenamefont {{Okawa}},\ and\ \citenamefont {{Pani}}}]{2014PhRvD..90d4069C}%
  \BibitemOpen
  \bibfield  {author} {\bibinfo {author} {\bibfnamefont {V.}~\bibnamefont
  {{Cardoso}}}, \bibinfo {author} {\bibfnamefont {L.~C.~B.}\ \bibnamefont
  {{Crispino}}}, \bibinfo {author} {\bibfnamefont {C.~F.~B.}\ \bibnamefont
  {{Macedo}}}, \bibinfo {author} {\bibfnamefont {H.}~\bibnamefont {{Okawa}}}, \
  and\ \bibinfo {author} {\bibfnamefont {P.}~\bibnamefont {{Pani}}},\ }\href
  {\doibase 10.1103/PhysRevD.90.044069} {\bibfield  {journal} {\bibinfo
  {journal} {\prd}\ }\textbf {\bibinfo {volume} {90}},\ \bibinfo {eid} {044069}
  (\bibinfo {year} {2014})},\ \Eprint {http://arxiv.org/abs/1406.5510}
  {arXiv:1406.5510 [gr-qc]} \BibitemShut {NoStop}%
\bibitem [{\citenamefont {{Chirenti}}\ and\ \citenamefont
  {{Rezzolla}}(2016)}]{2016PhRvD..94h4016C}%
  \BibitemOpen
  \bibfield  {author} {\bibinfo {author} {\bibfnamefont {C.}~\bibnamefont
  {{Chirenti}}}\ and\ \bibinfo {author} {\bibfnamefont {L.}~\bibnamefont
  {{Rezzolla}}},\ }\href {\doibase 10.1103/PhysRevD.94.084016} {\bibfield
  {journal} {\bibinfo  {journal} {\prd}\ }\textbf {\bibinfo {volume} {94}},\
  \bibinfo {eid} {084016} (\bibinfo {year} {2016})},\ \Eprint
  {http://arxiv.org/abs/1602.08759} {arXiv:1602.08759 [gr-qc]} \BibitemShut
  {NoStop}%
\bibitem [{\citenamefont {Yang}\ \emph {et~al.}(2017)\citenamefont {Yang},
  \citenamefont {Yagi}, \citenamefont {Blackman}, \citenamefont {Lehner},
  \citenamefont {Paschalidis}, \citenamefont {Pretorius},\ and\ \citenamefont
  {Yunes}}]{2017arXiv170105808Y}%
  \BibitemOpen
  \bibfield  {author} {\bibinfo {author} {\bibfnamefont {H.}~\bibnamefont
  {Yang}}, \bibinfo {author} {\bibfnamefont {K.}~\bibnamefont {Yagi}}, \bibinfo
  {author} {\bibfnamefont {J.}~\bibnamefont {Blackman}}, \bibinfo {author}
  {\bibfnamefont {L.}~\bibnamefont {Lehner}}, \bibinfo {author} {\bibfnamefont
  {V.}~\bibnamefont {Paschalidis}}, \bibinfo {author} {\bibfnamefont
  {F.}~\bibnamefont {Pretorius}}, \ and\ \bibinfo {author} {\bibfnamefont
  {N.}~\bibnamefont {Yunes}},\ }\href {\doibase 10.1103/PhysRevLett.118.161101}
  {\bibfield  {journal} {\bibinfo  {journal} {Phys. Rev. Lett.}\ }\textbf
  {\bibinfo {volume} {118}},\ \bibinfo {pages} {161101} (\bibinfo {year}
  {2017})}\BibitemShut {NoStop}%
\bibitem [{\citenamefont {{Maselli}}\ \emph
  {et~al.}(2017{\natexlab{a}})\citenamefont {{Maselli}}, \citenamefont
  {{Kokkotas}},\ and\ \citenamefont {{Laguna}}}]{2017arXiv170201110M}%
  \BibitemOpen
  \bibfield  {author} {\bibinfo {author} {\bibfnamefont {A.}~\bibnamefont
  {{Maselli}}}, \bibinfo {author} {\bibfnamefont {K.~D.}\ \bibnamefont
  {{Kokkotas}}}, \ and\ \bibinfo {author} {\bibfnamefont {P.}~\bibnamefont
  {{Laguna}}},\ }\href {\doibase 10.1103/PhysRevD.95.104026} {\bibfield
  {journal} {\bibinfo  {journal} {\prd}\ }\textbf {\bibinfo {volume} {95}},\
  \bibinfo {eid} {104026} (\bibinfo {year} {2017}{\natexlab{a}})},\ \Eprint
  {http://arxiv.org/abs/1702.01110} {arXiv:1702.01110 [gr-qc]} \BibitemShut
  {NoStop}%
\bibitem [{\citenamefont {{Sathyaprakash}}\ \emph {et~al.}(2012)\citenamefont
  {{Sathyaprakash}}, \citenamefont {{Abernathy}}, \citenamefont {{Acernese}},
  \citenamefont {{Ajith}}, \citenamefont {{Allen}}, \citenamefont
  {{Amaro-Seoane}}, \citenamefont {{Andersson}}, \citenamefont {{Aoudia}},
  \citenamefont {{Arun}}, \citenamefont {{Astone}},\ and\ \citenamefont
  {et~al.}}]{2012CQGra..29l4013S}%
  \BibitemOpen
  \bibfield  {author} {\bibinfo {author} {\bibfnamefont {B.}~\bibnamefont
  {{Sathyaprakash}}}, \bibinfo {author} {\bibfnamefont {M.}~\bibnamefont
  {{Abernathy}}}, \bibinfo {author} {\bibfnamefont {F.}~\bibnamefont
  {{Acernese}}}, \bibinfo {author} {\bibfnamefont {P.}~\bibnamefont {{Ajith}}},
  \bibinfo {author} {\bibfnamefont {B.}~\bibnamefont {{Allen}}}, \bibinfo
  {author} {\bibfnamefont {P.}~\bibnamefont {{Amaro-Seoane}}}, \bibinfo
  {author} {\bibfnamefont {N.}~\bibnamefont {{Andersson}}}, \bibinfo {author}
  {\bibfnamefont {S.}~\bibnamefont {{Aoudia}}}, \bibinfo {author}
  {\bibfnamefont {K.}~\bibnamefont {{Arun}}}, \bibinfo {author} {\bibfnamefont
  {P.}~\bibnamefont {{Astone}}}, \ and\ \bibinfo {author} {\bibnamefont
  {et~al.}},\ }\href {\doibase 10.1088/0264-9381/29/12/124013} {\bibfield
  {journal} {\bibinfo  {journal} {Classical and Quantum Gravity}\ }\textbf
  {\bibinfo {volume} {29}},\ \bibinfo {eid} {124013} (\bibinfo {year}
  {2012})},\ \Eprint {http://arxiv.org/abs/1206.0331} {arXiv:1206.0331 [gr-qc]}
  \BibitemShut {NoStop}%
\bibitem [{\citenamefont {Berti}\ \emph {et~al.}(2006)\citenamefont {Berti},
  \citenamefont {Cardoso},\ and\ \citenamefont {Will}}]{PhysRevD.73.064030}%
  \BibitemOpen
  \bibfield  {author} {\bibinfo {author} {\bibfnamefont {E.}~\bibnamefont
  {Berti}}, \bibinfo {author} {\bibfnamefont {V.}~\bibnamefont {Cardoso}}, \
  and\ \bibinfo {author} {\bibfnamefont {C.~M.}\ \bibnamefont {Will}},\ }\href
  {\doibase 10.1103/PhysRevD.73.064030} {\bibfield  {journal} {\bibinfo
  {journal} {Phys. Rev. D}\ }\textbf {\bibinfo {volume} {73}},\ \bibinfo
  {pages} {064030} (\bibinfo {year} {2006})}\BibitemShut {NoStop}%
\bibitem [{\citenamefont {Berti}\ \emph {et~al.}(2016)\citenamefont {Berti},
  \citenamefont {Sesana}, \citenamefont {Barausse}, \citenamefont {Cardoso},\
  and\ \citenamefont {Belczynski}}]{PhysRevLett.117.101102}%
  \BibitemOpen
  \bibfield  {author} {\bibinfo {author} {\bibfnamefont {E.}~\bibnamefont
  {Berti}}, \bibinfo {author} {\bibfnamefont {A.}~\bibnamefont {Sesana}},
  \bibinfo {author} {\bibfnamefont {E.}~\bibnamefont {Barausse}}, \bibinfo
  {author} {\bibfnamefont {V.}~\bibnamefont {Cardoso}}, \ and\ \bibinfo
  {author} {\bibfnamefont {K.}~\bibnamefont {Belczynski}},\ }\href {\doibase
  10.1103/PhysRevLett.117.101102} {\bibfield  {journal} {\bibinfo  {journal}
  {{Phys. Rev. Lett.}}\ }\textbf {\bibinfo {volume} {117}},\ \bibinfo {pages}
  {101102} (\bibinfo {year} {2016})}\BibitemShut {NoStop}%
\bibitem [{\citenamefont {{Maselli}}\ \emph
  {et~al.}(2017{\natexlab{b}})\citenamefont {{Maselli}}, \citenamefont
  {{V{\"o}lkel}},\ and\ \citenamefont {{Kokkotas}}}]{paper3}%
  \BibitemOpen
  \bibfield  {author} {\bibinfo {author} {\bibfnamefont {A.}~\bibnamefont
  {{Maselli}}}, \bibinfo {author} {\bibfnamefont {S.~H.}\ \bibnamefont
  {{V{\"o}lkel}}}, \ and\ \bibinfo {author} {\bibfnamefont {K.~D.}\
  \bibnamefont {{Kokkotas}}},\ }\href@noop {} {\bibfield  {journal} {\bibinfo
  {journal} {ArXiv e-prints}\ } (\bibinfo {year} {2017}{\natexlab{b}})},\
  \Eprint {http://arxiv.org/abs/1708.02217} {arXiv:1708.02217 [gr-qc]}
  \BibitemShut {NoStop}%
\bibitem [{\citenamefont {{Maggio}}\ \emph {et~al.}(2017)\citenamefont
  {{Maggio}}, \citenamefont {{Pani}},\ and\ \citenamefont
  {{Ferrari}}}]{2017arXiv170303696M}%
  \BibitemOpen
  \bibfield  {author} {\bibinfo {author} {\bibfnamefont {E.}~\bibnamefont
  {{Maggio}}}, \bibinfo {author} {\bibfnamefont {P.}~\bibnamefont {{Pani}}}, \
  and\ \bibinfo {author} {\bibfnamefont {V.}~\bibnamefont {{Ferrari}}},\
  }\href@noop {} {\bibfield  {journal} {\bibinfo  {journal} {ArXiv e-prints}\ }
  (\bibinfo {year} {2017})},\ \Eprint {http://arxiv.org/abs/1703.03696}
  {arXiv:1703.03696 [gr-qc]} \BibitemShut {NoStop}%
\bibitem [{\citenamefont {{Hod}}(2017)}]{2017arXiv170405856H}%
  \BibitemOpen
  \bibfield  {author} {\bibinfo {author} {\bibfnamefont {S.}~\bibnamefont
  {{Hod}}},\ }\href {\doibase 10.1007/JHEP06(2017)132} {\bibfield  {journal}
  {\bibinfo  {journal} {Journal of High Energy Physics}\ }\textbf {\bibinfo
  {volume} {6}},\ \bibinfo {eid} {132} (\bibinfo {year} {2017})},\ \Eprint
  {http://arxiv.org/abs/1704.05856} {arXiv:1704.05856 [hep-th]} \BibitemShut
  {NoStop}%
\bibitem [{\citenamefont {{Allen}}\ \emph {et~al.}(1998)\citenamefont
  {{Allen}}, \citenamefont {{Andersson}}, \citenamefont {{Kokkotas}},\ and\
  \citenamefont {{Schutz}}}]{1998PhRvD..58l4012A}%
  \BibitemOpen
  \bibfield  {author} {\bibinfo {author} {\bibfnamefont {G.}~\bibnamefont
  {{Allen}}}, \bibinfo {author} {\bibfnamefont {N.}~\bibnamefont
  {{Andersson}}}, \bibinfo {author} {\bibfnamefont {K.~D.}\ \bibnamefont
  {{Kokkotas}}}, \ and\ \bibinfo {author} {\bibfnamefont {B.~F.}\ \bibnamefont
  {{Schutz}}},\ }\href {\doibase 10.1103/PhysRevD.58.124012} {\bibfield
  {journal} {\bibinfo  {journal} {\prd}\ }\textbf {\bibinfo {volume} {58}},\
  \bibinfo {eid} {124012} (\bibinfo {year} {1998})},\ \Eprint
  {http://arxiv.org/abs/gr-qc/9704023} {gr-qc/9704023} \BibitemShut {NoStop}%
\end{thebibliography}%
%
%
\section{Appendix}\label{Appendix}
%
%
\subsection{Notes about the Implementation}\label{Notes about the Implementation}
%
The quasi-normal mode spectrum of the trapped modes is a discrete set of complex numbers which has to be inter- and extrapolated in the integration schemes. For all systems we studied, the spectrum is well-behaved and allows for such a treatment. If not mentioned differently, we use splines of degree three where interpolation is required.
\par
The maximum of the potential barrier $E_\text{max}$ is taken from the Regge-Wheeler potential for the corresponding harmonic index $l$. For the transmission $T(E)$ we use again splines to interpolate between the first and last trapped mode. Furthermore, we assume that it extrapolates to $1$ at $E_\text{max}$ and is equal to zero for $E=0$. The last assumption comes from the fact that the integral in equation \eqref{trasmission1} diverges for $E\rightarrow 0$, because the Regge-Wheeler potential falls off like $1/{r^{*}}^2$ for large $r^{*}$. Since the transmission grows exponentially, we use the logarithmic spline interpolation.
\par
Gravastars have an exponentially small $E_\text{min}>0$, but it can happen that the standard extrapolation $n(E_\text{min})+1/2=0$ yields $E_\text{min} < 0$. This is only due to the specific choice of extrapolation and we assume a very small, but positive value for $E_\text{min}=1e-5$. The influence of this choice  for larger values of $E$ is negligible. Once again we note that the results for very small $E$ should not be trusted.  
A negative region in the potential signals the presence of instabilities thus we excluded such cases in the present work. Even though, for the reconstruction of the width $\pazocal{L}_1(E)$ negative values are in general no problem for the method. However, if the external potential has the Regge-Wheeler form, one would not be able to find a unique solution for $V(r^{*})< 0$, because there is no third turning point for $E<0$. Since most potentials of alternative models do not have negative regions, we do not consider such cases here.
\par
For $\pazocal{L}_2(E)$ we additionally take into account the behavior around the minimum and close to the top of the barrier. For the values $E=[E_\text{min}, E_0]$ we use $\pazocal{L}_2(E) \approx a+b/\sqrt{E}$.  
{We choose this extrapolation because it takes into account that the Regge-Wheeler potential is proportional to $1/{r^{*}}^2$ for large values of $r^{*}$. This corresponds to small values of $E$ and provides the dominant contribution in the width of the Regge-Wheeler barrier. In this case the left side of the barrier falls off much faster than the right side (exponentially). The extrapolation is valid for small $E$ as long as $\pazocal{L}_2(E)$ is much larger as $\pazocal{L}_1(E)$.}
For the values $E=[E_N, E_\text{max}]$, where $E_N$ is the largest trapped mode, we use a simple parabola for $\pazocal{L}_2(E)$. In both cases the parameters are determined from the neighbouring trapped modes.
%
\subsection{Equivalence of Potentials}\label{EquivalentPotentials}
%
In order to reconstruct the potential one has to construct $\pazocal{L}_1(E)$ from the inverted spectrum $n(E)$. However, from the available data we only know the discrete spectrum $n(E_n)$. In order to use it in the integration methods one needs to use interpolation and extrapolation to create continuous expressions for $n(E)$ from $E_\text{min}$ to $E_\text{max}$. This step introduces, to some degree, a dependence on the specific techniques used. 
But, once the continuous spectrum $n(E)$ is constructed, the important integral in equation \eqref{transmission2} is the same for every potential that is reconstructed using  $\pazocal{L}_1(E)$.
The proof of this statement follows.
\par
The knowledge of $\pazocal{L}_1(E)$ alone, allows for the reconstruction of  infinitely many potentials. All of them yield the same spectrum for $E_n$ in the semi-classical approximation and therefore also the same discrete values for $n(E_n)$. The latter property implies that the Bohr-Sommerfeld rule 
\begin{equation}\label{appequi1}
\int_{x_0}^{x_1} \sqrt{E_n-V(x)} \mathrm{d}x  = \pi \left(n+\frac{1}{2}\right),
\end{equation}
is equivalent for any of these potentials.  Strictly speaking equation \eqref{appequi1} is valid for the discrete spectrum $E_n$, but a priori is not the same for all WKB spectrum equivalent potentials for continuous $E$ i.e. there might be ``fine structure'' details in the potential which cannot be traced by the semi-classical method.
However, since the inter- and extrapolated spectrum $n(E)$ is the same for all the potentials we can reconstruct from it, one finds that both sides of \eqref{appequi1} are also valid for continuous $E$ for all possible potentials.
\par
To prove that the approximate relation presented in equation \eqref{transmission2} is the same for all reconstructed potentials, we can study the symmetric potential $V_\text{S}(x)$ and a second arbitrary potential $\bar{V}(x)$. Both satisfy equation \eqref{appequi1} for continuous $E$ with their corresponding turning points $(x_{0\text{S}}, x_{1\text{S}})$ for the symmetric potential and $(\bar{x}_{0}, \bar{x}_{1})$ for the arbitrary potential. The validity of \eqref{appequi1} for continuous $E$ is used in the next step. Taking its absolute derivative with respect to $E$ and keeping in mind that each turning point implicitly also depends on $E$, it follows that
\begin{align}
\frac{\text{d}}{\text{d}E} \int_{x_0(E)}^{x_1(E)} \sqrt{E-V(x)} \mathrm{d}x
&=\left(\frac{\partial }{\partial E}+\frac{\text{d}x_0}{\text{d}E}\frac{\partial }{\partial x_0}+\frac{\text{d}x_1}{\text{d}E}\frac{\partial }{\partial x_1} \right)\int_{x_0(E)}^{x_1(E)} \sqrt{E-V(x)} \mathrm{d}x
\\
&=\frac{1}{2} \int_{x_0(E)}^{x_1(E)} \frac{1}{\sqrt{E-V(x)}} \text{d} x - \frac{\text{d}x_0}{\text{d}E} \sqrt{E-V(x_0(E))}+ \frac{\text{d}x_1}{\text{d}E} \sqrt{E-V(x_1(E))}
\\
&=\frac{1}{2} \int_{x_0(E)}^{x_1(E)} \frac{1}{\sqrt{E-V(x)}} \text{d} x,\label{appequi2}
\end{align}
where in the last step the elementary definition of turning points was used ($E-V(x)=0$). It now follows that all potentials which fulfill \eqref{appequi1} satisfy \eqref{appequi2} and therefore also for discrete $E_n$
\begin{align}
\int_{x_{0\text{S}}(E_n)}^{x_{1\text{S}}(E_n)} \frac{1}{\sqrt{E_n-V_\text{S}(x)}} \text{d} x
=\int_{\bar{x}_0(E_n)}^{\bar{x}_1(E_n)} \frac{1}{\sqrt{E_n-\bar{V}(x)}} \text{d} x.
\end{align}
This proves that \eqref{transmission2} does not depend on the specific choice of the potential within the family of reconstructed potentials, which was constructed from the specific choice of the inter- and extrapolated $n(E)$.
%
\subsection{Additional Results}\label{Additional Results}
\begin{figure}[H]
\centering
	\begin{minipage}{5.2cm}
	\includegraphics[width=5.7cm]{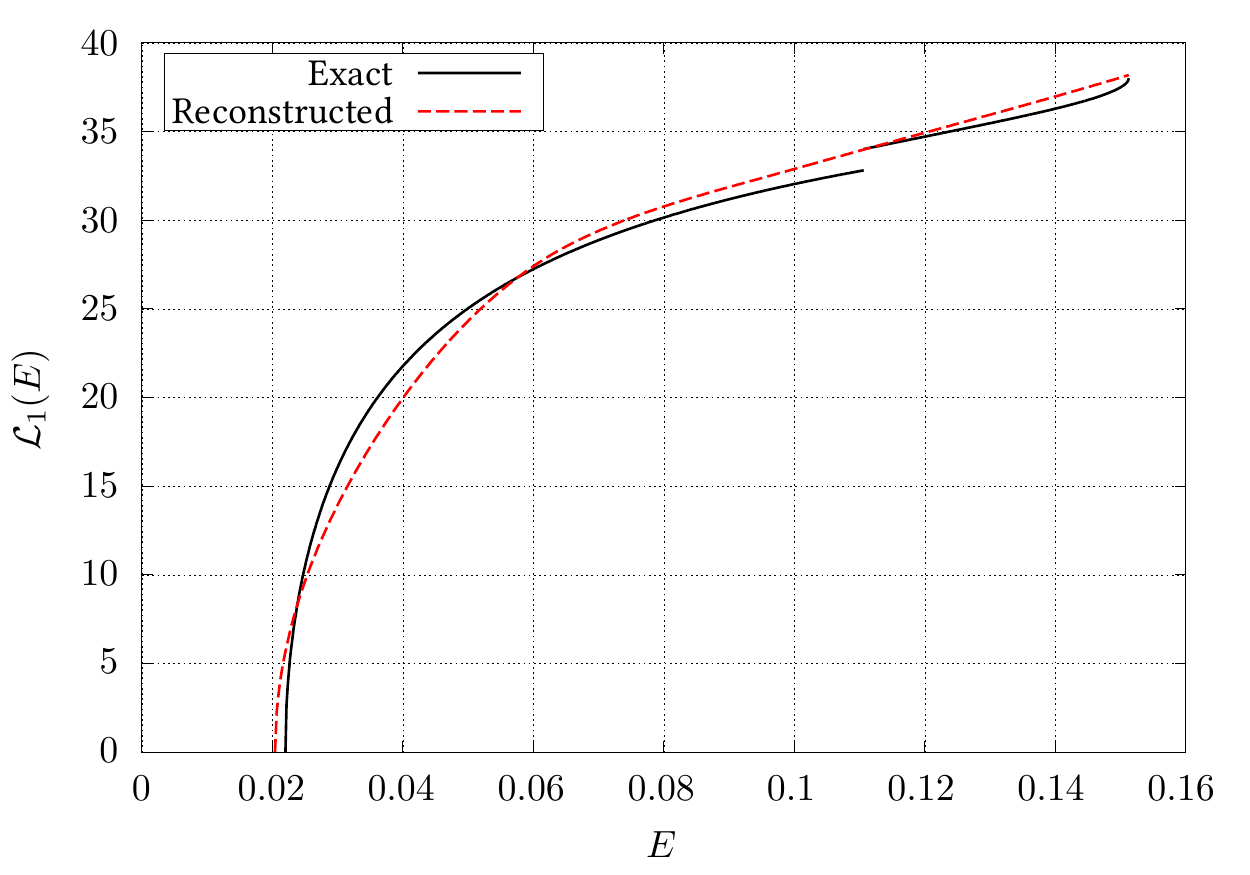}
	\end{minipage}
	\quad
	\begin{minipage}{5.2cm}
	\includegraphics[width=5.7cm]{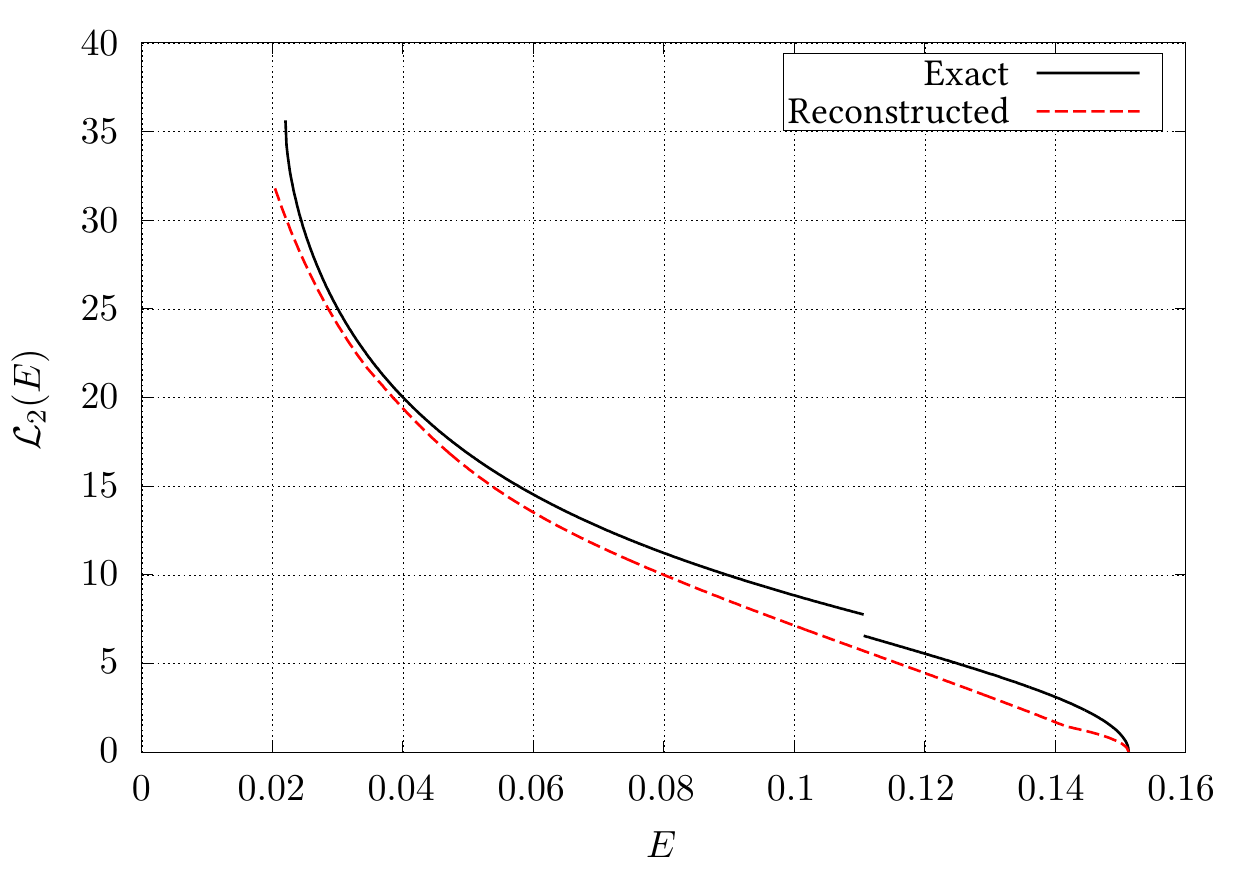}
	\end{minipage}
	\quad
	\begin{minipage}{5.2cm}
	\includegraphics[width=5.7cm]{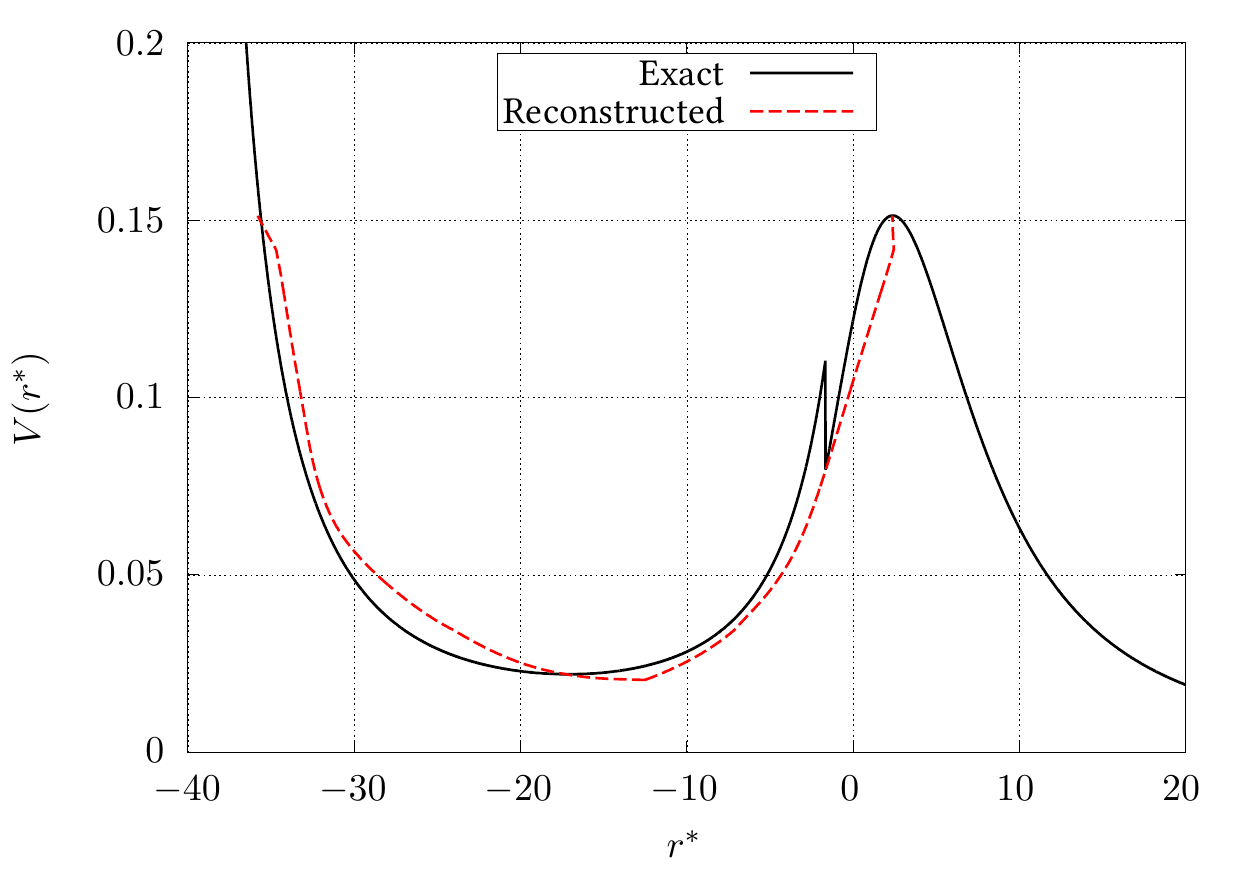}
	\end{minipage}
\caption{Constant density star with $R/M=2.28$ and $l=2$\label{CS_228_l2}, for this case there exist 4 trapped modes. \textbf{Left panel:} width of the bound region $\pazocal{L}_1(E)$ for the exact potential (black) and the reconstructed one (red dashed). \textbf{Central panel:} width of the potential barrier $\pazocal{L}_2(E)$ for the exact potential (black) and the reconstructed one (red dashed). \textbf{Right panel:} exact axial mode potential (black) vs the reconstructed one (red dashed).}
\end{figure}

\begin{figure}[H]
\centering
	\begin{minipage}{5.2cm}
	\includegraphics[width=5.7cm]{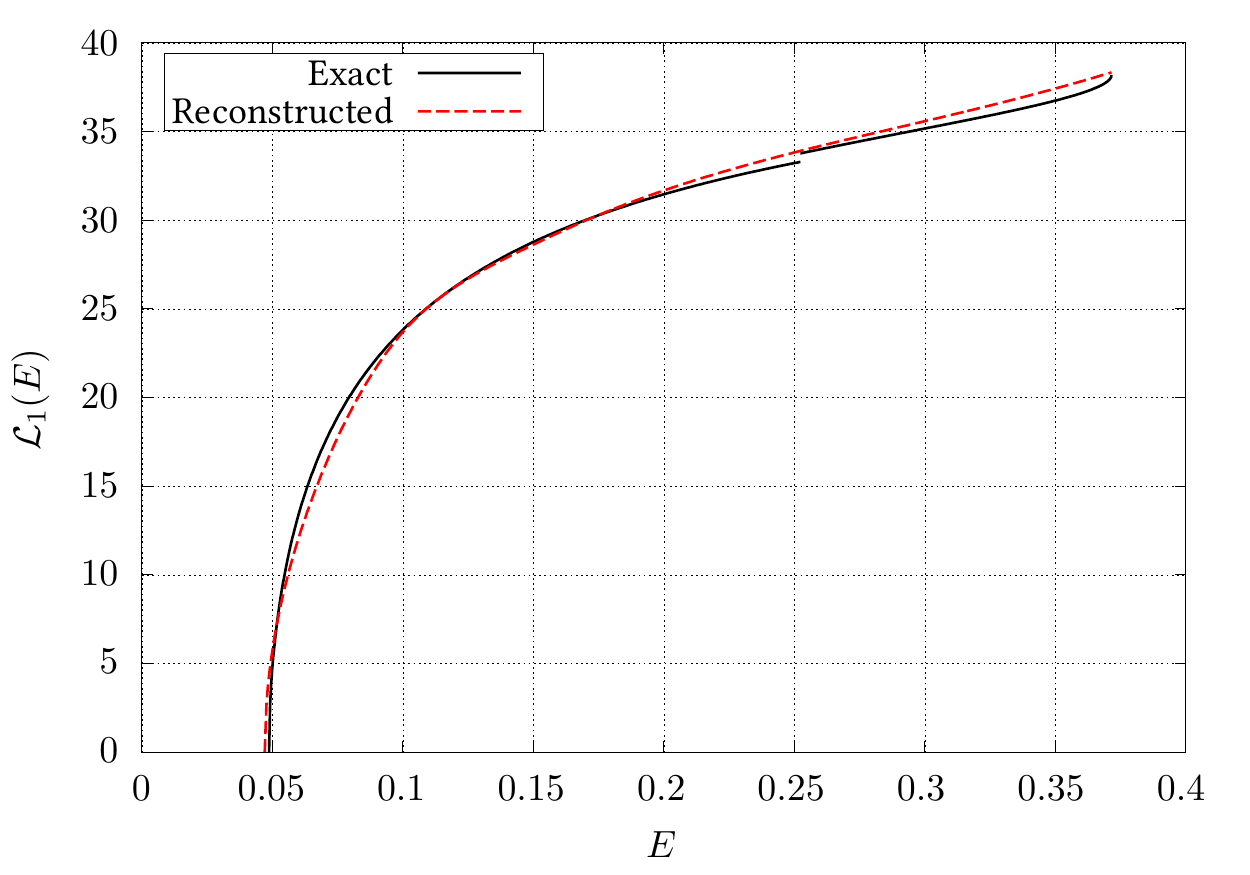}
	\end{minipage}
	\quad
	\begin{minipage}{5.2cm}
	\includegraphics[width=5.7cm]{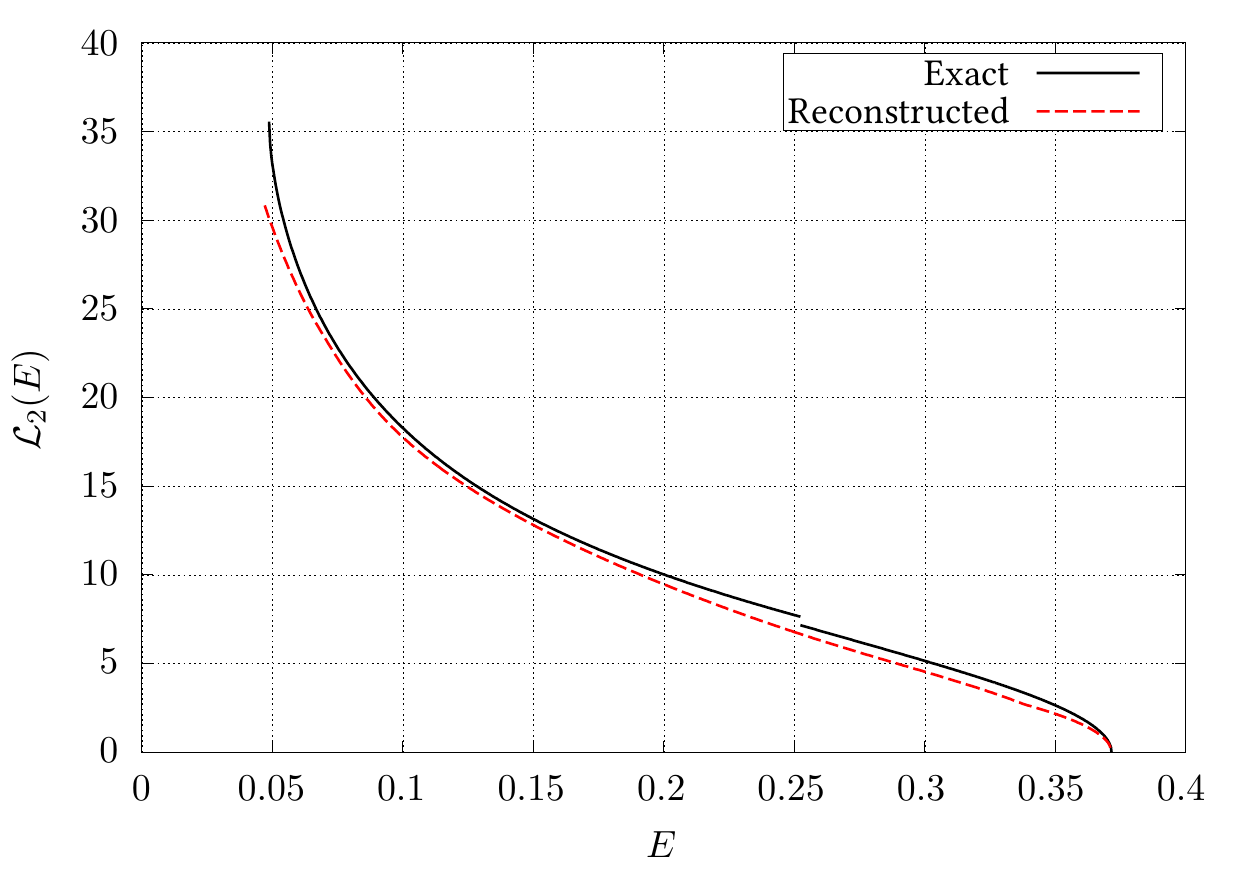}
	\end{minipage}
	\quad
	\begin{minipage}{5.2cm}
	\includegraphics[width=5.7cm]{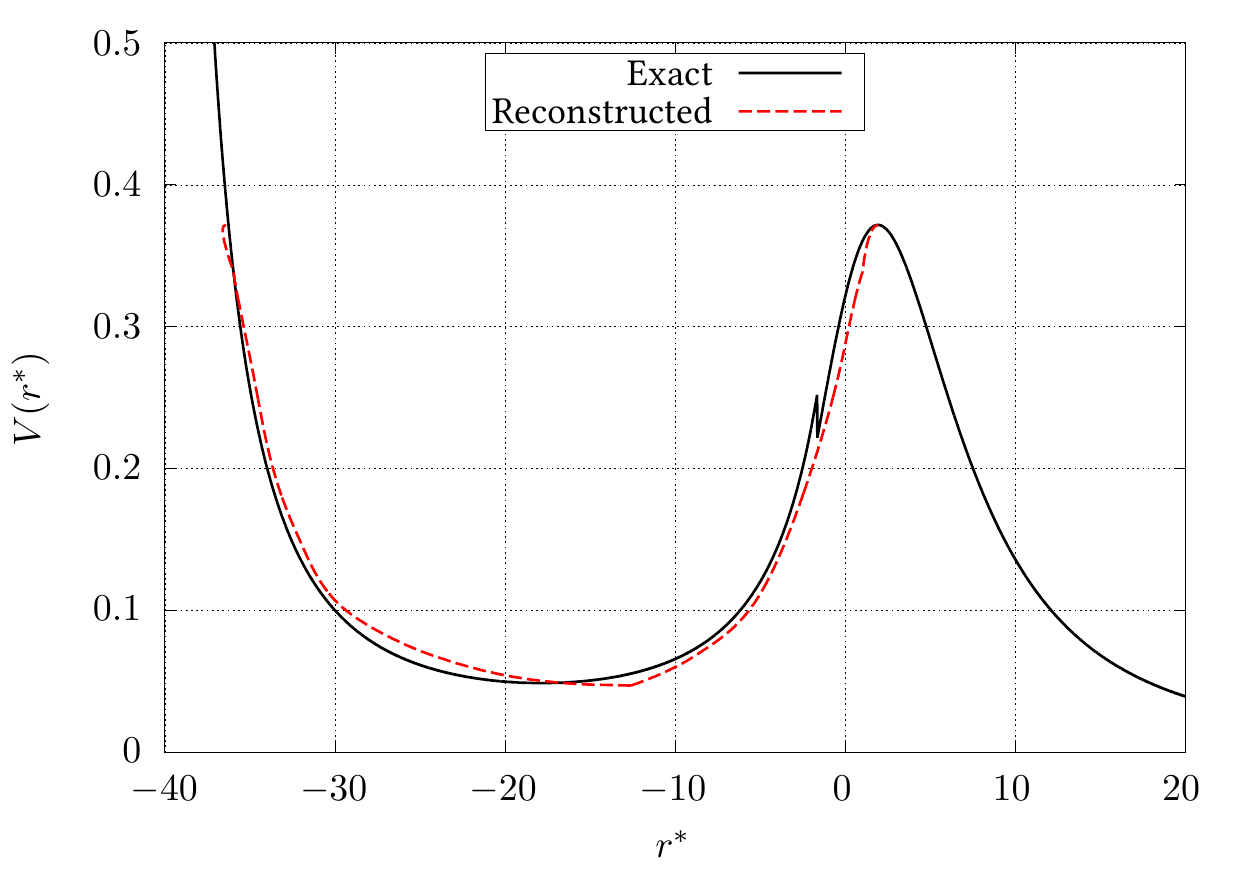}
	\end{minipage}
	\caption{Constant density star with $R/M=2.28$ and $l=3$ \label{CS_228_l3}, for this case there exist  6 trapped modes. \textbf{Left panel:} width of the bound region $\pazocal{L}_1(E)$ for the exact potential (black) and the reconstructed one (red dashed). \textbf{Central panel:} width of the potential barrier $\pazocal{L}_2(E)$ for the exact potential (black) and the reconstructed one (red dashed). \textbf{Right panel:} exact axial mode potential (black) vs the reconstructed one (red dashed).}
\end{figure}

\begin{figure}[H]
\centering
	\begin{minipage}{5.2cm}
	\includegraphics[width=5.7cm]{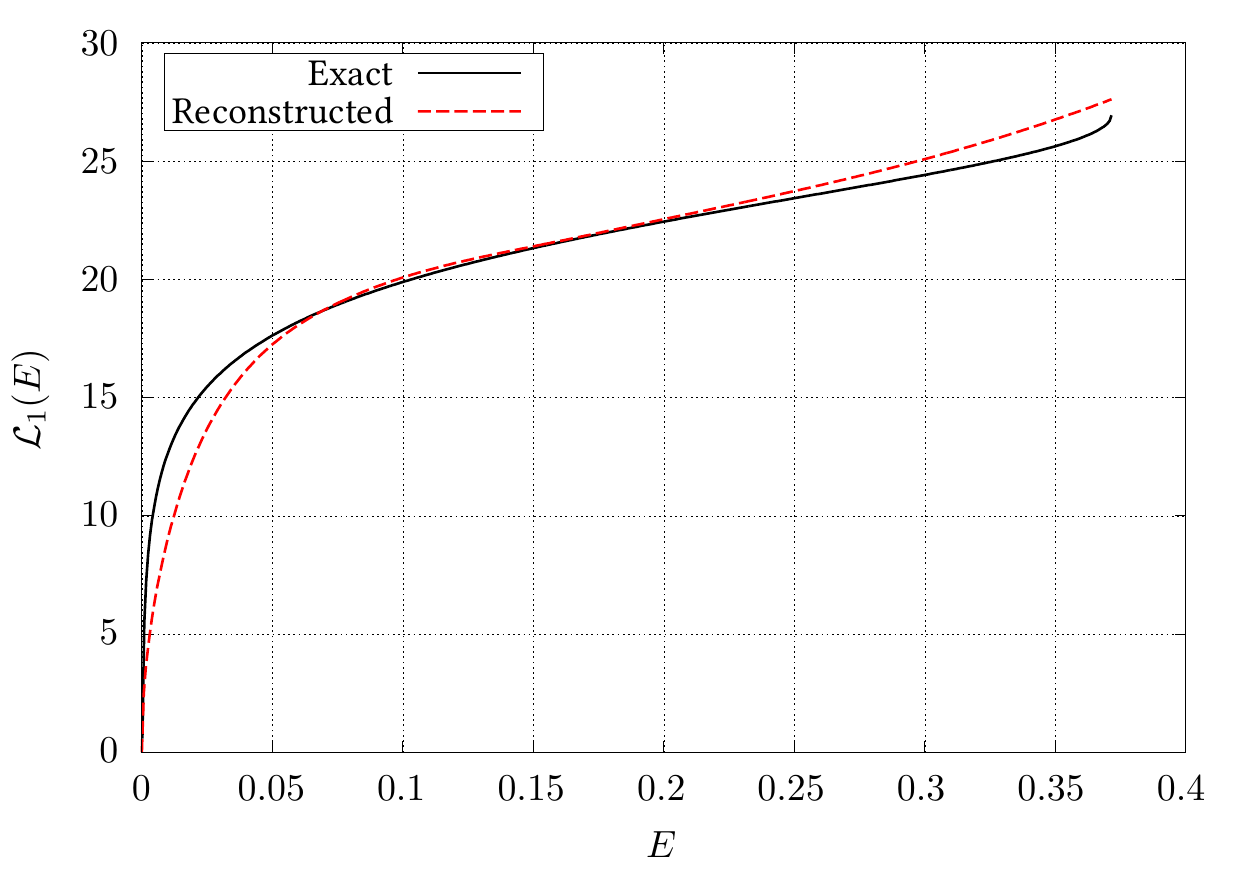}
	\end{minipage}
	\quad
	\begin{minipage}{5.2cm}
	\includegraphics[width=5.7cm]{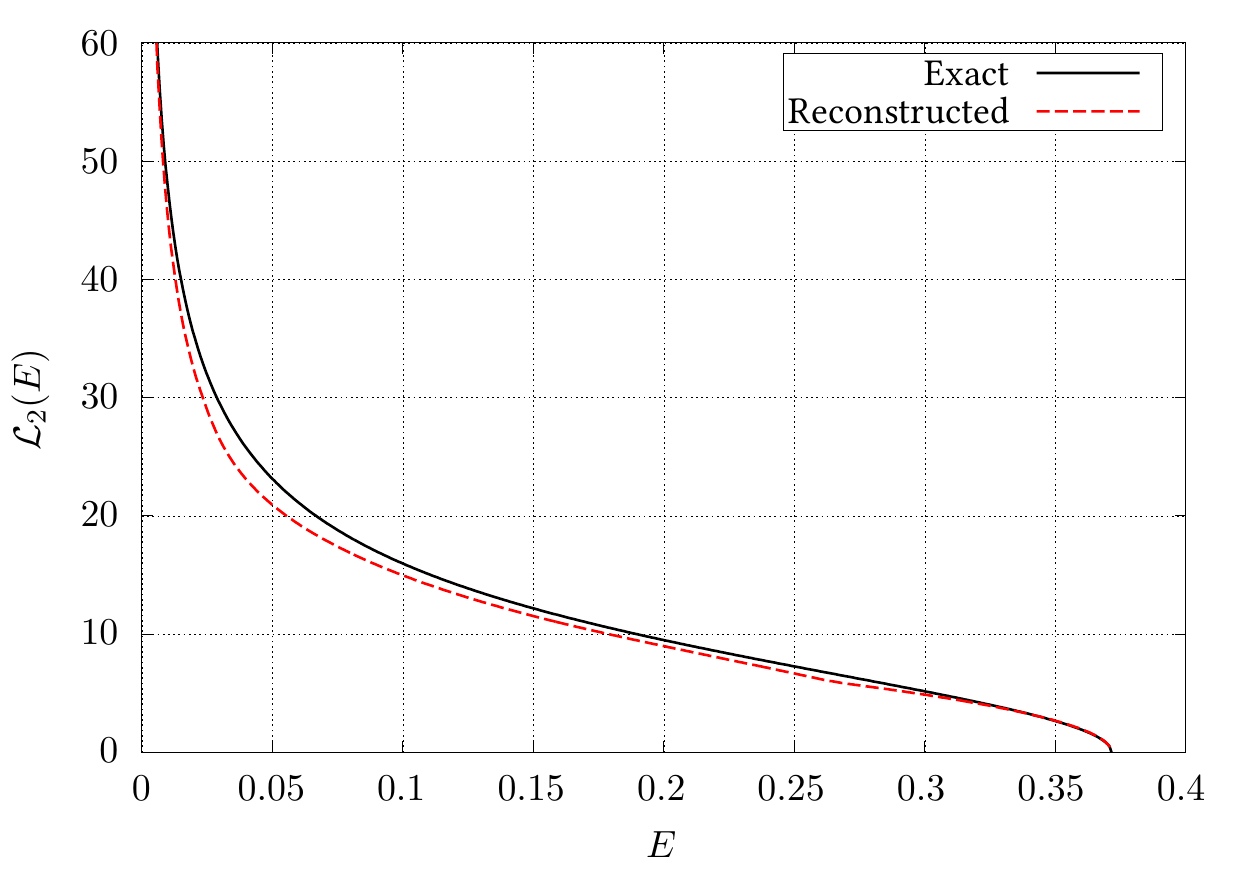}
	\end{minipage}
	\quad
	\begin{minipage}{5.2cm}
	\includegraphics[width=5.7cm]{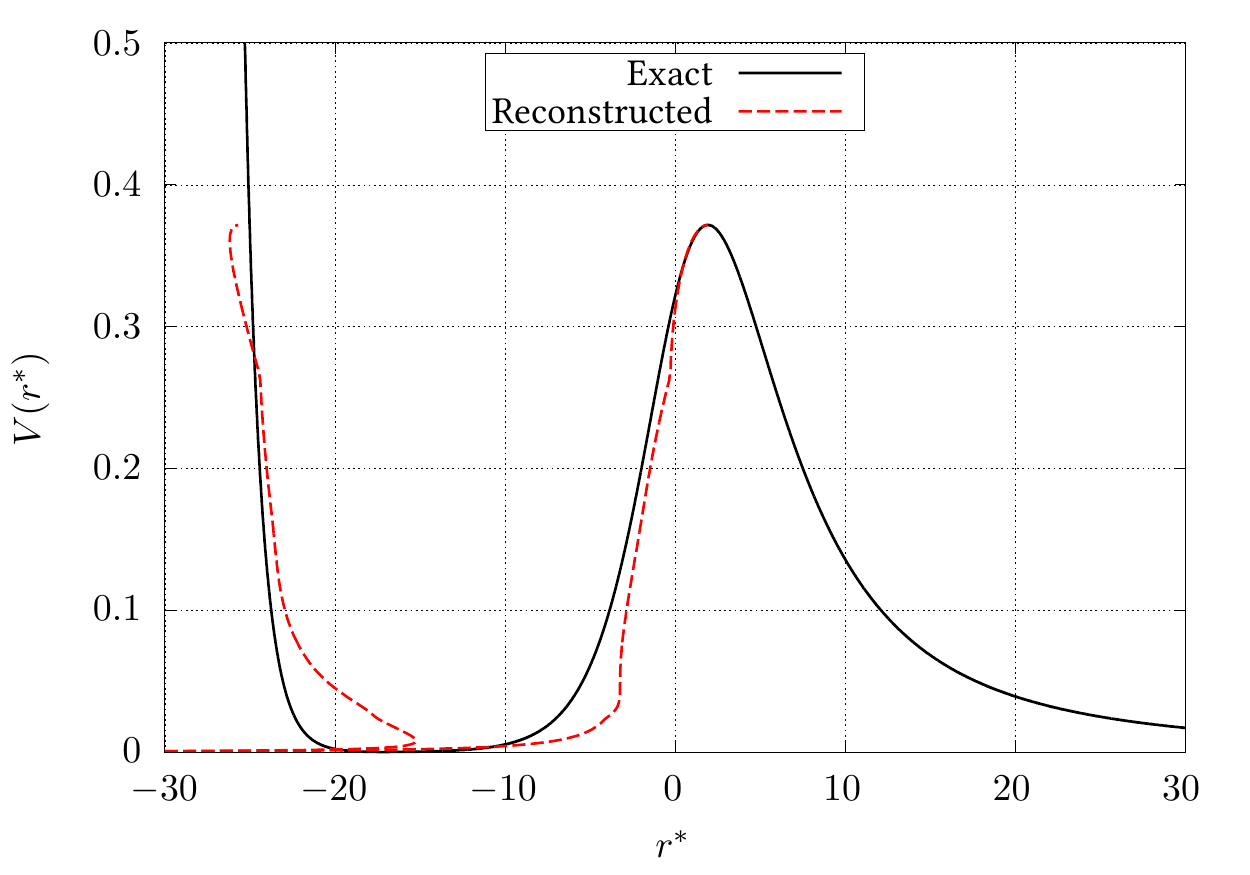}
	\end{minipage}
	\caption{Gravastar with $\mu=0.49997$ and $l=3$\label{GS_499970_l3}, for this case there exist  4 trapped modes. \textbf{Left panel:} width of the bound region $\pazocal{L}_1(E)$ for the exact potential (black) and the reconstructed one (red dashed). \textbf{Central panel:} width of the potential barrier $\pazocal{L}_2(E)$ for the exact potential (black) and the reconstructed one (red dashed). \textbf{Right panel:} exact axial mode potential (black) vs the reconstructed one (red dashed).}
\end{figure}

\begin{figure}[H]
	\centering
	\begin{minipage}{5.2cm}
		\includegraphics[width=5.7cm]{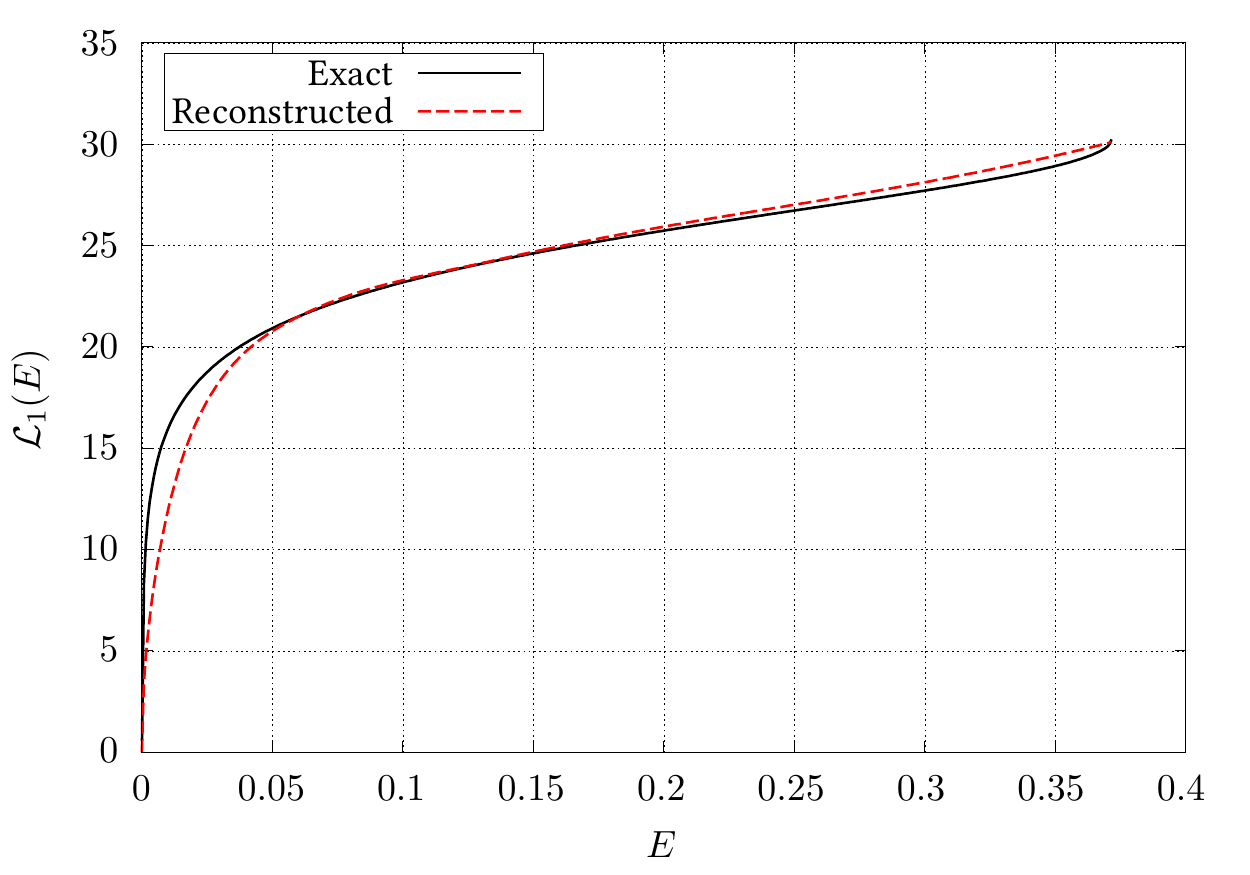}
	\end{minipage}
	\quad
	\begin{minipage}{5.2cm}
		\includegraphics[width=5.7cm]{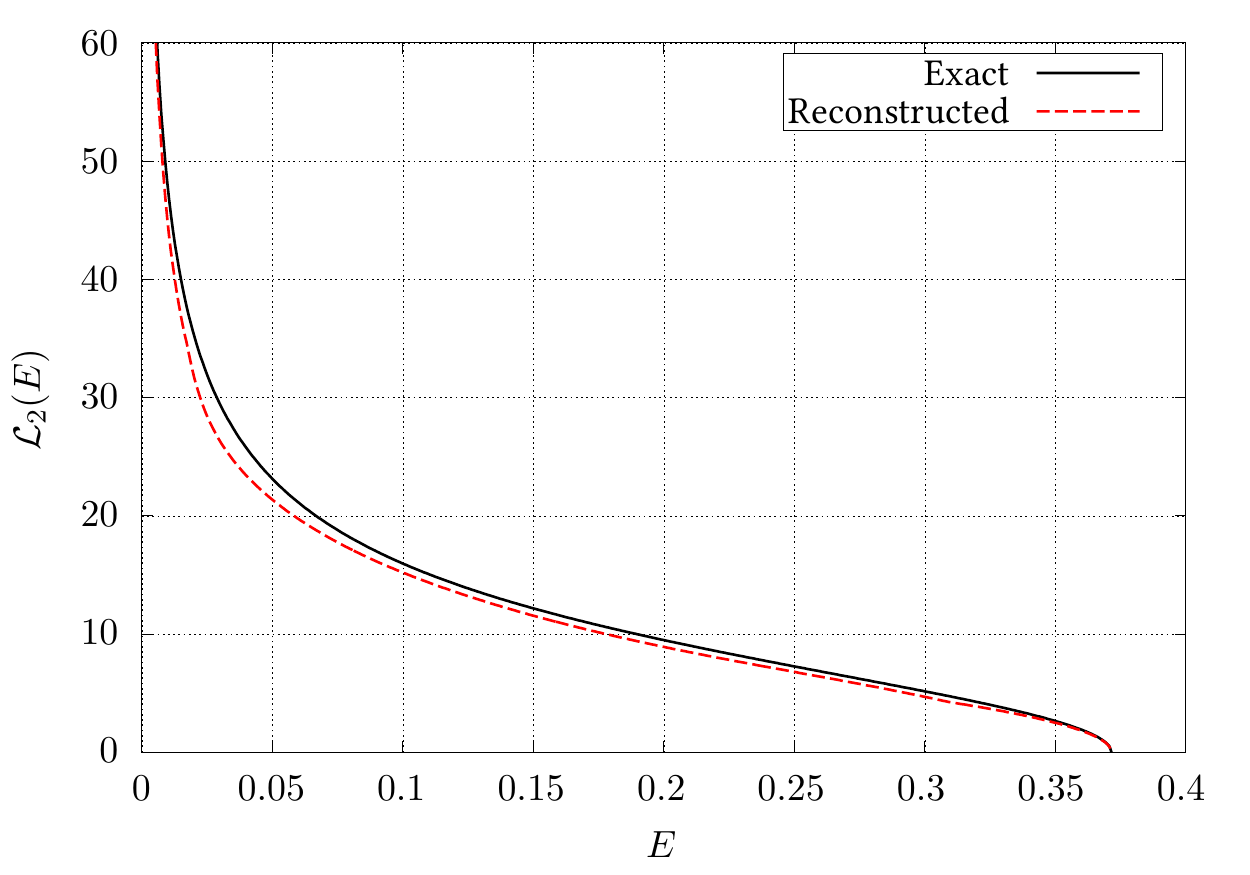}
	\end{minipage}
	\quad
	\begin{minipage}{5.2cm}
		\includegraphics[width=5.7cm]{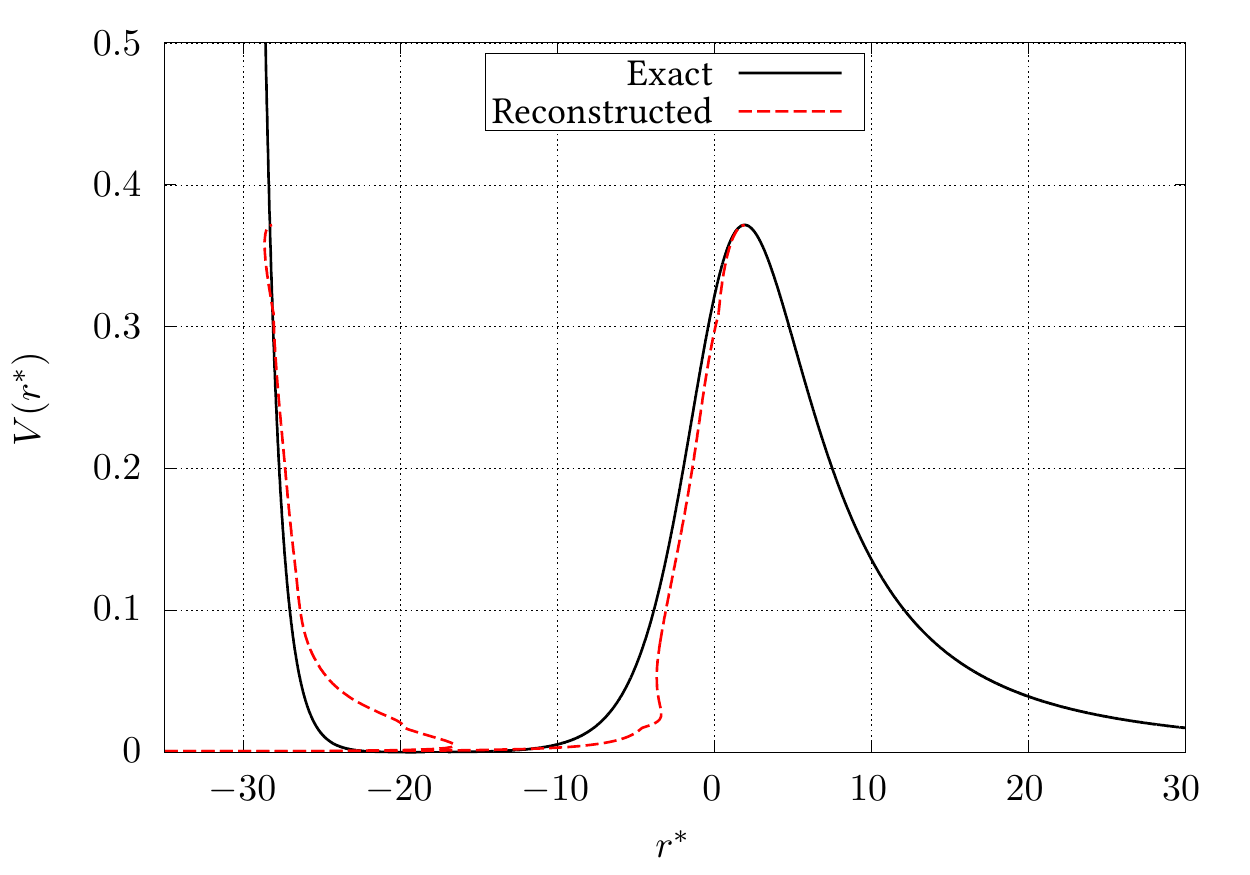}
	\end{minipage}
	\caption{Gravastar with $\mu=0.49999$ and $l=3$\label{GS_499990_l3}, for this case there exist 5 trapped modes. \textbf{Left panel:} width of the bound region $\pazocal{L}_1(E)$ for the exact potential (black) and the reconstructed one (red dashed). \textbf{Central panel:} width of the potential barrier $\pazocal{L}_2(E)$ for the exact potential (black) and the reconstructed one (red dashed). \textbf{Right panel:} exact axial mode potential (black) vs the reconstructed one (red dashed).}
\end{figure}

%
We provide additional results for constant density stars with $R/M=2.28$ in figures \ref{CS_228_l2} and  \ref{CS_228_l3}, as well as for gravastars with $\mu=0.49997$ and $\mu=0.49999$ in figures \ref{GS_499970_l3} and \ref{GS_499990_l3}.
%
\subsection{Tables for Trapped Modes}\label{Tables for Quasi-normal Modes}
%
In Table \ref{Table1} and Table \ref{Table3} we show the numerical values for the trapped quasi-normal modes used as input for the calculation of $\pazocal{L}_1(E)$ and $\pazocal{L}_2(E)$. They were produced by a code discussed in \cite{1994MNRAS.268.1015K}
\begin{table}[H]
\centering
\caption{The trapped axial modes for constant density stars with different $R/M$ and $l$. 
\label{Table1}}
\begin{tabular}{|c|ll|ll|ll|ll|}
\hline
 & \multicolumn{4}{|c|}{$R/M=2.26$} & \multicolumn{4}{|c|}{$R/M=2.28$} \\
\hline
 & \multicolumn{2}{|c|}{$l=2$} & \multicolumn{2}{|c|}{$l=3$} & \multicolumn{2}{|c|}{$l=2$} & \multicolumn{2}{|c|}{$l=3$}  \\
\hline
n & \text{Re}($\omega_n$) & \text{Im}($\omega_n$) & \text{Re}($\omega_n$) & \text{Im}($\omega_n$)& \text{Re}($\omega_n$) & \text{Im}($\omega_n$) & \text{Re}($\omega_n$) & \text{Im}($\omega_n$) \\
\hline
0	&	0.1090	&	1.240e-09	&	0.1508	&	1.520e-13	&	0.1856	&	6.200e-07	&	0.2568	&	9.630e-10	\\
1	&	0.1484	&	3.950e-08	&	0.1901	&	7.760e-12	&	0.2519	&	2.660e-05	&	0.3237	&	6.060e-08	\\
2	&	0.1876	&	5.470e-07	&	0.2293	&	1.620e-10	&	0.3160	&	4.630e-04	&	0.3902	&	1.640e-06	\\
3	&	0.2267	&	4.850e-06	&	0.2686	&	2.080e-09	&	0.3764	&	3.640e-03	&	0.4559	&	2.730e-05	\\
4	&	0.2654	&	3.230e-05	&	0.3078	&	1.930e-08	&		&		&	0.5201	&	3.080e-04	\\
5	&	0.3036	&	1.720e-04	&	0.3469	&	1.420e-07	&		&		&	0.5816	&	2.180e-03	\\
6	&	0.3410	&	7.300e-04	&	0.3860	&	8.790e-07	&		&		&		&		\\
7	&	0.3777	&	2.300e-03	&	0.4249	&	4.720e-06	&		&		&		&		\\
8	&		&		&	0.4636	&	2.250e-05	&		&		&		&		\\
9	&		&		&	0.5019	&	9.640e-05	&		&		&		&		\\
10	&		&		&	0.5397	&	3.630e-04	&		&		&		&		\\
11	&		&		&	0.5768	&	1.150e-03	&		&		&		&		\\
\hline
\end{tabular}
\end{table}
\begin{table}[H]
	\centering
	\caption{The trapped axial modes for gravastars for different values of $\mu$ and $l=3$. 
		\label{Table3}}
	\begin{tabular}{|c| ll| ll| ll|}
		\hline
		& \multicolumn{2}{|c|}{$\mu=0.49997$} & \multicolumn{2}{|c|}{$\mu=0.49999$} & \multicolumn{2}{|c|}{$\mu=0.499999$} \\
			\hline
			n & \text{Re}($\omega_n$) & \text{Im}($\omega_n$) & \text{Re}($\omega_n$) & \text{Im}($\omega_n$)& \text{Re}($\omega_n$) & \text{Im}($\omega_n$)  \\
			\hline
0	&	0.1500	&	3.8550e-11	&	0.1304	&	8.3806e-12	&	0.1019	&	6.7921e-13	\\
1	&	0.2848	&	5.2885e-08	&	0.2508	&	9.5436e-09	&	0.1994	&	5.3054e-10	\\
2	&	0.4059	&	7.7891e-06	&	0.3614	&	1.2144e-06	&	0.2916	&	5.0750e-08	\\
3	&	0.5149	&	3.9039e-04	&	0.4635	&	5.8802e-05	&	0.3790	&	1.9883e-06	\\
4	&		&		&	0.5562	&	1.3301e-03	&	0.4620	&	4.5774e-05	\\
5	&		&		&		&		&	0.5398	&	6.6507e-04	\\
			\hline 
		\end{tabular}
	\end{table}
\nocite{}
\end{document}